\documentclass[10.5pt,oneside]{report}

\usepackage[utf8]{inputenc}
\usepackage[T1]{fontenc}
\usepackage{lmodern}
\usepackage[a4paper,margin=1.2in]{geometry}
\usepackage{microtype}
\usepackage{amsmath,amssymb,amsthm,mathrsfs}
\usepackage{graphicx}
\usepackage{subcaption}
\usepackage{tabularx}
\usepackage{booktabs}
\usepackage{longtable}
\usepackage{array}
\usepackage{ragged2e}
\usepackage{enumitem}
\usepackage{threeparttable}
\usepackage{multirow}
\usepackage{algorithm}
\usepackage{algorithmic}
\usepackage{siunitx}
\usepackage{setspace}
\usepackage{etoolbox}
\usepackage{chngcntr}
\usepackage[natbibapa]{apacite}
\usepackage[hidelinks]{hyperref}

\graphicspath{{figures/}}
\makeatletter
\@removefromreset{figure}{chapter}
\@removefromreset{table}{chapter}
\@removefromreset{equation}{chapter}
\makeatother

\makeatletter
\renewcommand\footnotesize{\@setfontsize\footnotesize{8pt}{11pt}}
\makeatother
\AtBeginEnvironment{longtable}{\fontsize{8.5pt}{10pt}\selectfont}
\counterwithout{footnote}{chapter}

\makeatletter

\renewcommand{\listoffigures}{%
  \section*{List of Figures}
  \addcontentsline{toc}{section}{List of Figures}
  \@starttoc{lof}
}

\renewcommand{\listoftables}{%
  \section*{List of Tables}
  \addcontentsline{toc}{section}{List of Tables}
  \@starttoc{lot}
}

\makeatother

\makeatletter
\renewcommand\normalsize{%
  \@setfontsize\normalsize{11.5pt}{14pt}%
  \abovedisplayskip 10pt plus 2pt minus 5pt%
  \belowdisplayskip \abovedisplayskip
}
\makeatother

\normalsize

\begin{document}

\begin{titlepage}
  \centering
  \vspace*{0.7in}

  {\sffamily\bfseries\LARGE
  Toward a Threat Actor Profiling Taxonomy for Pre-Release Risk Management of Open-Weight Frontier Models\par}

  \vspace{1.25in}

  {\large Practical Achievement submitted to\par}
  \vspace{0.12in}
  {\large\bfseries Tsinghua University\par}
  \vspace{0.12in}
  {\large in partial fulfillment of the requirement\par}
  {\large for the professional degree of\par}
  \vspace{0.12in}
  {\large\sffamily\bfseries Master of Global Affairs\par}

  {\large by\par}
  \vspace{0.12in}
  {\large\sffamily\bfseries James Q Zhang\par}

  \vspace{0.5in}
  {\large Supervisor: Professor Xue Lan\par}

    \vspace{0.5in}

  \begin{minipage}{0.72\textwidth}
    \centering\itshape\small
    This paper was originally submitted as a capstone Practical Achievement for the Master of Global Affairs degree at Schwarzman College, Tsinghua University in June 2026. This version has been lightly edited for upload to arXiv.
  \end{minipage}

  \vspace{0.45in}
  {\large\bfseries June 2026\par}
  \vspace{0.25in}
\end{titlepage}

\pagenumbering{roman}
\chapter*{Abstract}
\addcontentsline{toc}{chapter}{Abstract}

Pre-release risk management for frontier AI misuse risks routinely leaves threat actor assumptions implicit, inconsistently specified, or ungrounded. This capstone argues that explicit adversary characterization should be regarded as a prerequisite for evaluations that are interpretable, comparable, and faithful to the risks they target. We propose a six-attribute taxonomy (covering technical sophistication, prior domain knowledge, organizational capacity, operational infrastructure, financial capacity, and time horizon) with empirically grounded tiers derived from existing terrorism, biosecurity, and cybersecurity literature. The taxonomy is designed to function as research infrastructure: a common language for pre-specifying adversary assumptions before evaluations are conducted, analogous to pre-analysis plans for randomized controlled trials (RCTs) in medicine and economics. Its application is particularly urgent for open-weight model developers, for whom release decisions are irreversible and must anticipate adversarial reasoning.

\vspace{1em}
\noindent\textbf{Keywords:} frontier AI misuse; risk management; open-weight models; adversary profiling; capability evaluation

\tableofcontents
\newpage
\listoffigures
\listoftables

\clearpage
\pagenumbering{arabic}
\setcounter{page}{1}



\chapter{Introduction}


Open-weight AI models, whose development have historically lagged behind their closed counterparts, are rapidly closing the gap.  In some domains, open-weight model capabilities seem to already match or surpass those of proprietary models. At its release, Moonshot AI's Kimi K2.5 (January 2026) achieved state-of-the-art performances for agentic searching and browsing capabilities; DeepSeek's DeepSeek-V3.2-Speciale (December 2025) scored higher than OpenAI's GPT-5 and on par with Google's Gemini-3.0-Pro on many well-known benchmarks testing model reasoning \citep{kimi-k25,deepseek-32}. Had DeepSeek-V3.2-Speciale competed in the 2025 International Mathematical Olympiad (IMO), the most prestigious and challenging international math competition for high school students, it would have received a gold medal -- a feat only the top $\sim$10\% of scores achieve \citep{imo-gold-stat}. Troublingly, these capabilities also include those that are dual-use, which can be leveraged for chemical, biological, radiological, nuclear (CBRN) and offensive cyber misuse \citep{estimating-worst-case-risk}. As open-weight models continue to improve and even start to set what is considered frontier, such misuse implications grow more profound for how we think and act about the risks of releasing powerful model weights into the world without restriction\footnote{We acknowledge that the open-weight versus closed distinction is reductive. Model access exists on a spectrum, from fully closed APIs to conditionally gated releases to unrestricted weight downloads. Factors like documentation, licensing terms, and safety mitigation further complicate a binary perspective. Nevertheless, the decision of whether to release models weights to the general public remains an ongoing, concrete choice that AI model developers face, and we believe it is worth examining on such terms.}.

The future of the open-weight model ecosystem is particularly concerning because once weights are released, they can be used and modified without any authoritative oversight. Although this dynamic brings benefits such as research accessibility and decentralization of power, it also means ceding control: motivated malicious actors can disable external monitoring and filtering safeguards, fine-tune away internal safety behavior, or introduce harmful capabilities -- all under a cloak of secrecy \citep{open-tech-problems-ow-ai-risk-management}.

Chief among these concerns is the offense-defense balance problem in the realm of cybersecurity. The release of open-weight models is often defended on the grounds that broad access can enable better collective defense, as more researchers can audit systems, diversify safety research, and jointly identify vulnerabilities. Yet, it is becoming increasingly unclear whether this argument will stand \citep{cnas-cyber-balance, cset-cyber-balance}. Attackers just need one successful exploit, while defenders must guard against all possible threat vectors. This asymmetry constrains defenders' margin for error -- and that margin shrinks even further as the reliability deficits of existing AI tools prevent defenders from fully trusting and deploying them in the very systems most likely to be targeted. Early expert polling suggests a growing consensus that the balance may favor offense, at least in the short term \citep{ai-impact-cyber-balance}. 

Because open-weight models are structurally afforded less protection, favor attackers in the offense-defense balance, and cannot be recalled, pre-release practices face considerable pressure to get it right. Safety benchmarking via automated assessments, expert red-teaming, and uplift trials has become the dominant procedure for evaluating models before deployment, and it has produced genuine insight into model capabilities \citep{metr-common-elements}. However, this approach has a critical blind spot. As David Krueger, a professor of machine learning at Mila, has noted in a February 2026 address to the Canadian House of Commons, ``[t]he kinds of tests we have can show that an AI system is dangerous. They cannot show that it is safe." Looking ahead, he stresses that ``[w]e should also not expect any amount of investment to solve this problem in the foreseeable future" \citep{krueger-canada}. 

This sentiment has already reverberated in industry. In the same month as David Krueger's address, the major AI developer Anthropic acknowledged that they could not confidently rule out whether their newest publicly available model, Claude Opus 4.6, could assist ``moderately-resourced state programs" with developing weapons of mass destruction, even after extensive evaluations. They caution: ``a clear rule-out of the next capability threshold may soon be difficult or impossible under the current regime" \citep{opus-46-model-card}. If evaluations are better at raising alarms than providing the assurance needed, how do we know when the risks of releasing models open-weight are too high?

Answering that question may require tighter coupling of capability-based evaluations with adversary-focused frameworks, which are well-established in cybersecurity tradition, that seek full understanding of attacker strategies, objectives, and constraints rather than model capabilities in isolation \citep{gdm-ai-cyber-framework, kaloudi-ai-cyber-landscape}. Existing approaches such as the Cyber Kill Chain \citep{cyber-attack-chain-lockheed} and the MITRE ATT\&CK framework (extended to AI systems through MITRE ATLAS\footnote{See https://atlas.mitre.org/matrices/ATLAS.}) \citep{mitre-attack} have served as valuable infrastructure for modeling how threat actors operate, but they are organized around observed behavior and built retrospectively. Ex-ante adversary\footnote{For simplicity, we use the term \textit{adversary} and \textit{threat actor} interchangeably.} profiling, structured around specific attributes of potential threat actors, could strengthen risk analysis in ways that neither capability evaluations nor post-hoc behavioral taxonomies currently offer. 

This paper argues that attribute-based adversary profiling should become a standardized component of pre-release practice alongside safety benchmarking for open-weight model developers\footnote{We frame our suggestion around open-weight models because it presents a particularly severe case, given the irreversibility of weight release and lack of post-deployment safeguards. Ideally, the practice should nonetheless be considered a critical component of any responsible pre-release procedure, regardless of deployment modality.}. Namely, by employing a bottom-up framework that deconstructs threat actors into various key attributes, researchers can set up evaluations that better align with the risks they are concerned about – and thereby  produce risk analyses that are more accurate, clearly scoped, and interoperable across the field. Our contribution is thus twofold: introducing a six-attribute taxonomy to systematize the characterization of potential threat actors, and demonstrating how applying this taxonomy helps developers reason more precisely about risks and design more targeted evaluations. 

\section{Scope}
Our analysis focuses on the two frontier misuse threats most prominently flagged in the literature -- particularly. chemical, biological, radiological, nuclear (CBRN) and offensive cyber \citep{international-ai-safety-report-2026}. Broad governance questions, such as staged release policy and determining acceptable risk levels, remain important open questions we leave for future work. This paper instead offers a readily usable taxonomy and its operationalization for pre-release workflows to assist with consequential decisions in AI deployment.

\section{Methodology}
Synthesizing published research to develop an analytical approach, our research design proceeds in three stages. First, we conduct a literature review of the current state of frontier AI misuse risk management, identifying where both practitioners and researchers have noted gaps. Second, we formulate the threat actor attributes based on existing analysis on cybersecurity and CBRN threats, while making sure they are distinct and independently meaningful \citep{taxonomy-dev}. For each attribute, we distinguish multiple tiers by drawing from well-established theories from terrorism studies and cybersecurity; where literature is sparse, we rely on empirical data sources to ground an attribute's tier system in observable and verifiable criteria. Lastly, we provide recommendations on how each attribute can inform evaluation design and how this taxonomy-based classification system can become live practice.



\chapter{Background}

This chapter details the technical, institutional, and conceptual landscape within which our paper intervenes. Section \ref{sec:2-open-weight-ai-background} foregrounds the concern of increasingly powerful open-weight models, highlighting why they warrant particular attention. Section \ref{sec:2-current-risk-practices} surveys the risk management practices that inform model release deliberations. Section \ref{sec:2-evaluate-frontier-risk} gives an overview of how frontier misuse risks are evaluated today. Section \ref{sec:2-dev-currently-characterization} reviews how major developers already explicitly characterize the threat actors they are concerned about. 

\section{The Open-Weight Model Problem}
\label{sec:2-open-weight-ai-background}

\begin{figure}[!htbp]
  \centering
  \includegraphics[width=0.8\linewidth]{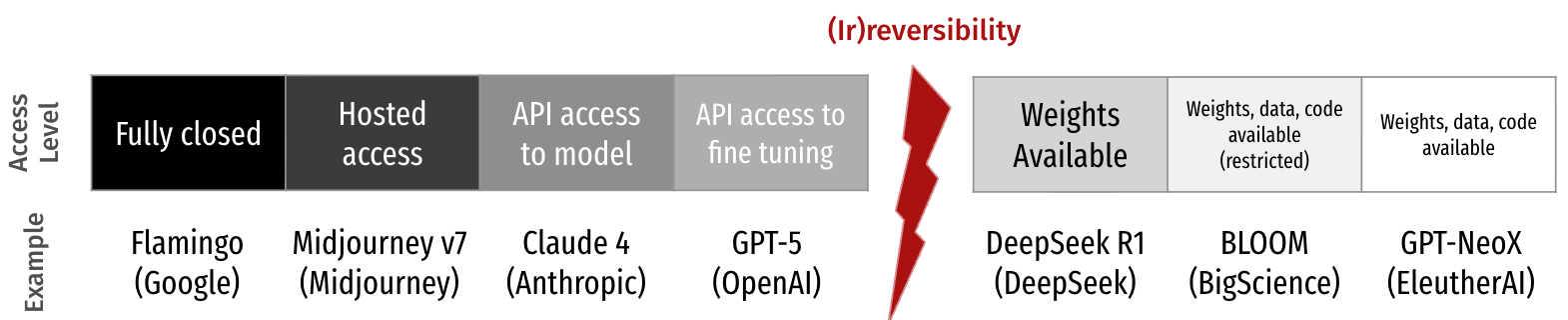}
  \caption[Model Access Spectrum]{Models deployed in manners left of the red marker are usually referred to as ``closed." Models deployed in formats to the marker's right, known as open-weight models, carry the irreversibility dilemma. Source: adapted from \citep{international-ai-safety-report-2026, considerations-for-governing-open-foundation-models}.}
  \label{fig:OAI_spectrum}
\end{figure}

\textbf{The Openness Spectrum.}  Model access exists on a spectrum, ranging from full closed systems reserved for a company's internal use through various intermediate points (e.g., hosted access, API access for inference) to fully open releases where model weights, data, and code are all available without restriction. The critical boundary for our paper rests at the point where model weights become publicly available. Figure \ref{fig:OAI_spectrum} shows this gradient and marks the transition with a red break. Models whose weights are publicly downloadable exhibit five distinctive properties: broader access, greater customizability, local adaptation and inference ability, inability to rescind model access, inability to monitor or moderate model usage \citep{societal-impact-open-models}. 

This boundary is significant because it marks a major qualitative shift in developers' relationships to their models after release. Releasing with an access level left of the boundary in Figure \ref{fig:OAI_spectrum}, developers retain a degree of operational control. They can update safety filters, revoke API access, monitor usage patterns, or shut the system down entirely \citep{benchmark-misuse, constitutional-classifiers, anthropic-disrupt-misuse}. Releasing with access levels on the right, none of these interventions exist. Once users download model weights, the model can be copied, modified, fine-tuned, hosted independently, or redistributed in ways beyond the original developers' visibility, influence, or intent. In this sense, open-weight releases are irreversible, and are what make every pre-release action all the more important. 

\textbf{Approaching the Frontier Intensifies Judgment Calls.} The urgency of the open-release question has heightened because the capability gap between open-weight and frontier closed models has narrowed considerably. The AI governance research institute Epoch AI has estimated the lag to be between six months to a year, a figure that the 2026 International AI Safety Report reiterated \citep{epoch-year-lag, international-ai-safety-report-2026}. This convergence draws concern because the skills in biology, chemistry, code generation that make frontier models commercially and academically valuable seem to also be the ones that make them dangerous in the wrong hands. When assessing the tamper resistance of their new open-weight model series prior to release in August 2025, OpenAI found  their gpt-oss-120b model performed just 1.2 percentage points worse than their frontier closed o3 model on biorisk tacit knowledge and troubleshooting; other leading open-weight models at time, such as DeepSeek's DeepSeek R1-0528 and Moonshot AI's Kimi K2, performed comparably. Notably, the best models on this task, which included both open-weight and closed models, scored within five percentage points away from their estimate of human expert performance. The report concluded that releasing gpt-oss could add new biorisk capabilities \citep{estimating-worst-case-risk}.

Governmental bodies have formally acknowledged restricting public frontier open-weight releases as one of the strongest possible safeguards, but no clear escalation mechanisms exist for acting upon it \citep{aisi-managing-risks-ow, ntia-report}. Excluding the competitive and geopolitical pressures that disincentivize restraint from a single developer, acting on this option also requires a degree of evaluative confidence that existing methods have yet to provide \citep{international-ai-safety-report-2026}.



\section{Current Risk Management Practices}
\label{sec:2-current-risk-practices}

The 2026 International AI Safety Report, the product of a collaboration spanning more than 100 independent experts and 30 countries, breaks down risk management for general-purpose AI into four components: identification, analysis, mitigation, and governance \citep{international-ai-safety-report-2026}. Risk identification entails efforts that detail the nature and scope of potential harms. To accomplish this, companies and research institutions have widely turned towards threat modeling exercises and risk taxonomy development. The former describes the structured process of fleshing out the pathways through which a threat may undergo to realize harm. Among major AI developers, common threat models include their products helping build biological weapons, carry out cyberattacks, and automate AI research itself \citep{microsft-fsp, openai-preparedness-framework}. Building risk taxonomies complements by organizing possible harms into categories -- MIT's AI Risk Repository, which brings together over 1,700 risks across the literature is a prominent example \footnote{This repository features three components: a risk database, a casual taxonomy, and a domain taxonomy. See https://airisk.mit.edu/ for more.}. Other ways for identifying risks include consulting with  experts and running bounty programs, which are initiatives set up to financially reward external parties helping developers uncover vulnerabilities \citep{openai-march-bug-bounty, gray-swan-2025-redteam}. 

Risk analysis involves determining the likelihood and severity of identified harms and whether further action is warranted. The dominant method at this stage is model capability evaluation -- systematic testing designed to gauge a model's ability in performing tasks relevant to a risk domain. To address the harms that could cause mass casualties or large-scale financial damages, many large AI companies have already voluntarily committed to conducting targeted evaluations \citep{metr-common-elements}. The specifics underlying these assessments vary greatly and are discussed at length in the following section. As an escalation mechanism that complements these risk measurement practices, companies have also begun specifying capability thresholds that, if surpassed, trigger additional scrutiny and safeguarding measures \citep{metr-common-elements, meta-fsp, anthropic-rsp-v3.1}. A central concept employed in many analyses is marginal risk, the degree to which introducing a given model increases risks beyond what is already within reach \citep{societal-impact-open-models}. Interpreting such results can be challenging and misleading, however, as comparisons require shared counterfactual baselines that are often difficult to establish \citep{international-ai-safety-report-2026}. 

Risk mitigation encompasses the actions that can be taken to bring identified risks down to acceptable levels. Given that no single safeguard has proven sufficient, the prevailing conceptual framework has relied on building defense-in-depth, commonly illustrated as stacked layers of Swiss cheese: while threats can seep through the holes of any one layer, having multiple layers of protection in series make the overall system more robust. The 2026 International AI Safety Report classifies measures into four layers: training interventions, deployment interventions, post-deployment monitoring, and societal resilience \citep{international-ai-safety-report-2026}. At the implementation level, model developers may also look to strengthen the security of model weights against theft, set guidelines for acceptable use, and impose user screening for access \footnote{For an example, see Claude's identity verification policy: https://support.claude.com/en/articles/14328960-identity-verification-on-claude} \citep{rand-secure-model-weights}. 

Risk governance connects these previous three components together, holding the overall risk management profile accountable. The primary instruments here are documentation and transparency practices. Model cards \footnote{Example cards for Anthropic's family of models can be found here: https://www.anthropic.com/system-cards}, usually published alongside a model's release, have become the main vehicle for disclosing that very model's capabilities, limitations, safety evaluations, and risk mitigation strategies. At the regulatory level, California's SB 53 law \footnote{See Section 22757.12(a): https://legiscan.com/CA/text/SB53/id/3270002.} requires frontier developers to proactively share their AI governance and risk mitigation practices; EU's AI Act similarly imposes documentation and transparency obligations \footnote{See Article 53: https://artificialintelligenceact.eu/article/53/.}. Pertaining to the CBRN and offensive cyber misuse risks that are within the scope of our paper, governmental involvement has thus far largely penetrated at the level of enforcing transparency rather than prescribing specific technical treatments \citep{international-ai-safety-report-2026}.  

Institutionalized at the AI Seoul Summit in May 2024, frontier safety policies (FSPs) have become a mainstream signal among AI companies for exemplifying responsible risk management \citep{seoul-commitments}. FSPs detail how companies themselves plan to evaluate, monitor, and mitigate severe AI harms. Structurally, they operate as nested if-then commitments: if a model's evaluated capabilities cross a defined threshold in a given risk domain, then specific, pre-defined safeguarding measures are triggered in response. For example, crossing a lower priority threshold may result in a company restricting model access to certain, verified users. Crossing a relatively high threshold may mean not releasing a model publicly at all, as was the case with Anthropic's Mythos model for cybersecurity concerns \footnote{See Anthropic's plans for its restricted Mythos model via Project Glasswing: https://www.anthropic.com/glasswing}. In nearly all of the published FSPs \footnote{As of April 21, 2026, based on the research nonprofit METR's FSP tracker: https://metr.org/fsp.}, triggers are set for CBRN and offense cyber capabilities \citep{metr-common-elements}. Despite their growing adoption, the real-world effectiveness of these frameworks remains largely untested, and it is unclear whether such commitments would hold under pressure \citep{international-ai-safety-report-2026, white-house-voluntary}.

\section{Evaluating Frontier AI Misuse Risks}
\label{sec:2-evaluate-frontier-risk}

Three principal methods are employed in misuse evaluations for CBRN and offensive cyber capabilities: automated assessments, expert red-teaming, and human uplift trials. Each encodes different assumptions and tradeoffs about the threat actor being modeled, with direct implications for what the resulting evidence can and can not plausibly support. Such work carries special importance for open-weight models, where the layer of deployment-time safeguards is largely absent and bare weights can be attacked with minimal restriction \citep{estimating-worst-case-risk}.

Automated assessments involve the model answering multiple-choice questions, responding to open-ended prompts, and completing multi-step agentic tasks. They are designed to allow for systematic capability measurement without heavy manual oversight and can be run a scale. Well-known examples include LAB-Bench (bioresearch ability) \citep{lab-bench}, WMDP (bio-, cyber-, and chemical- security knowledge) \citep{wmdp-benchmark}, Cybench (cyber-relevant tasks) \citep{cybench}, and SecureBio VCT (virology protocol troubleshooting) \citep{vct-benchmark}. In more detail, LAB-Bench tests models on providing literature reviews, interpreting figures, and manipulating DNA sequences using more than 2,400 multiple choices questions. The VCT evaluation presents 322 scenario-based questions developed by PhD-level virologists that test the know-how needed for operating in virology laboratories. Because these evaluations are automated, they afford reproducibility and breadth; however, they are limited in modeling the interactive, iterative nature of real-world misuse processes. 

Expert red-teaming engages domain experts to simulate adversaries, stress-testing a model to uncover weaknesses and elicit risky capabilities \citep{international-ai-safety-report-2026, safe-harbor}. The Anthropic Opus 4.6 Model Card \footnote{See Sections 8.2.4.6 and 8.2.4.7: 
https://www-cdn.anthropic.com/6a5fa276ac68b9aeb0c8b6af5fa36326e0e166dd.pdf.} and the OpenAI GPT-5 system card \footnote{See Sections 5.3.3.3 and 5.3.3.4: https://arxiv.org/pdf/2601.03267.} describe efforts of this kind. Testers can look to probe how well a model plans new experiments, distinguish between high and low value ideas, or serves as a lab assistant. An emerging belief from Anthropic's Opus 4.6 red-teaming exercises places it as a research force multiplier incapable of generating truly novel insights. Nevertheless, it is important to note that results from such studies are sensitive to the specific personnel involved and resources provided, making the research design -- and, in particular, the threat actor profile calibrated against -- especially consequential \citep{ai-rct-paper}.

Human uplift trials \footnote{This approach draws upon the randomized controlled trial (RCT) methodology well-established in medical and economic research \citep{ai-rct-paper}.} directly operationalize the concept of marginal uplift. Treatment groups with model access are compared against control groups without it, and performance differences are taken as evidence of the model's contribution to real-world risk \citep{ai-rct-paper}. A representative example comes from a March 2026 collaboration among Scale AI, SecureBio, the University of Oxford, and UC Berkeley, in which novice participants equipped with access to multiple frontier LLMs (such as OpenAI's o3, Anthropic's Claude Opus 4, and Google's Gemini Deep Research) were compared against a similarly novice population restricted to standard internet search on biosecurity knowledge tasks \citep{scaleai-uplift-test}. The LLM-assisted group performed approximately four times more accurately than their control counterparts. Because these studies are structurally configured to measure joint human-AI performance, uplift trials are increasingly cited in crucial deployment and governance decisions \citep{international-ai-safety-report-2026}. Like those from expert red-teaming exercises, their results are nonetheless sensitive to design choices (e.g., participant population composition, time duration, and task selection) -- and it is currently common for stakeholders to make judgment calls without a clear understanding of where a given study's assumptions fail to generalize \citep{ai-rct-paper}. 

Ultimately, the evaluation ecosystem remains methodologically immature, as assessments are frequently opaque in their implementation. While opacity may reduce infohazard risks, it complicates independent quality verification and the comparability of findings across different efforts \citep{actually-measure}. One underexplored opportunity within this landscape is a more systematic, fine-grained manner to characterize potential threat actors -- the gap our research seeks to address. Greater specificity in detailing who evaluations are designed to model would render their underlying assumptions more legible, and therefore make scrutiny and cross-system comparisons more meaningful. 

\section{How Developers Currently Characterize Threat Actors}
\label{sec:2-dev-currently-characterization}

To take stock of where industry practice currently stands, we survey the twelve frontier model developers that have published FSPs \footnote{As of April 21, 2026.}. For each company policy, we extract every instance where it describes the threat actors themselves in context of CBRN and offensive cyber misuse risk. Table \ref{tab:threat_actor_characterizations} presents these characterizations verbatim, sorted alphabetically by developer name and separated by risk domain. Notice the heterogeneity across policies: not only in which dimensions of an actor are highlighted, but also in how concretely -- or whether -- any actor is described at all.  



\begin{longtable}{>{\raggedright\arraybackslash}p{2.0cm}
                  >{\raggedright\arraybackslash}p{2.2cm}
                  >{\raggedright\arraybackslash}p{4.5cm}
                  >{\raggedright\arraybackslash}p{4.5cm}}

    \caption{Explicit Threat Actor Characterizations in Frontier AI Safety Policies}
    \label{tab:threat_actor_characterizations} \\
    \toprule
    \textbf{Developer} & \textbf{Policy (Date)} & \textbf{CBRN} & \textbf{Offensive Cyber} \\
    \midrule
    \endfirsthead

    \caption[]{Explicit Threat Actor Characterizations in Frontier Model Safety Policies (continued)} \\
    \toprule
    \textbf{Developer} & \textbf{Policy (Date)} & \textbf{CBRN} & \textbf{Offensive Cyber} \\
    \midrule
    \endhead

    \bottomrule
    \multicolumn{4}{r}{\footnotesize\textit{Continued on next page}} \\
    \endfoot

    \bottomrule
    \multicolumn{4}{p{14.5cm}}{%
        \footnotesize\textit{Note.} Developers are listed alphabetically. All
        quotations are reproduced verbatim from the cited documents and reference pages are noted; ellipses
        reflect omissions of surrounding context not relevant specifically to the
        characterization. Within the CBRN and WMD categories, most developers
        have focused their explicit characterizations on chemical and biological
        risks; coverage of radiological and nuclear threat actors are more limited in
        the surveyed policies. Developers listed as ``N/A'' either do not address threat actors in the relevant
        domain or discuss CBRN and offensive cyber risks without specifically detailing the threat 
        actors themselves.} \\
    \endlastfoot


    \cite{amazon-fsp}
        & Frontier Model Safety Framework \newline (Feb.\ 10, 2025)
        & \begin{itemize}[itemsep=8pt]
            \item ``\ldots a non-subject matter expert\ldots'' (2)
        \end{itemize}
        & \begin{itemize}[itemsep=8pt]
            \item ``\ldots a moderately skilled actor (e.g., an individual with
          undergraduate level understanding of offensive cyber activities or
          operations)\ldots'' (2)
        \end{itemize} \\
    \midrule

    \cite{anthropic-compliance-f, anthropic-rsp-v3.1}
        & Responsible Scaling Policy (RSP) v3.1 \newline (Apr.\ 2, 2026) \newline \newline \newline
          Frontier Compliance Framework (FCF) \newline (Mar.\ 2026)
        & \textit{Non-novel:} 

        \begin{itemize}[itemsep=8pt]
            \item ``\ldots individuals or groups with basic
          technical backgrounds (e.g., undergraduate STEM degrees)\ldots''
          (RSP 6; FCF 5)
          \item ``\ldots individual users and relatively
          small teams\ldots'' (RSP 6)
        \end{itemize}
        
        \hspace{10pt}
        
          \textit{Novel:} 

          \begin{itemize}[itemsep=8pt]
              \item ``\ldots threat actors (for example, moderately
          resourced expert-backed teams)\ldots'' (RSP 7; FCF 5)
          \item ``\ldots well-resourced and -staffed threat actors\ldots''
          (RSP 7)
          \item``\ldots strongest and most plausible threat actors
          that are not bound by a credible governance regime\ldots''
          (RSP 7)
          \end{itemize}
        & N/A \\
    \midrule

    \cite{cohere-fsp}
        & Secure AI Frontier Model Framework \newline (Feb.\ 7, 2025)
        & N/A
        & N/A \\
    \midrule

    \cite{g42-fsp}
        & Frontier AI Safety Framework \newline (Feb.\ 6, 2025)
        & \begin{itemize}[itemsep=8pt]
            \item ``\ldots prompt engineering, fine-tuning, and
          agentic tool usage\ldots'' (4--5)
          \item ``\ldots individual with
          only introductory biology experience\ldots'' (5)
          \item ``Even a determined actor\ldots'' (8)
        \end{itemize}
        & \begin{itemize}[itemsep=8pt]
            \item ``\ldots  prompt engineering, fine-tuning, and
          agentic tool usage\ldots'' (4--5)
          \item ``Even a determined actor\ldots'' (8)
        \end{itemize} \\
    \midrule

    \cite{gdm-fsf}
        & Frontier Safety Framework v3.0 \newline (Sep. 22, 2025)
        & \begin{itemize}[itemsep=8pt]
            \item ``\ldots low to medium resourced actors\ldots'' (10)
            
          \item ``\ldots the low to medium resourced actors who would be likely to
          experience the most CBRN uplift are unlikely to pose a substantial
          exfiltration threat at the level of RAND OC3 groups.'' (10)
        \end{itemize}
        & 
        \begin{itemize}[itemsep=8pt]
            \item ``\ldots well-resourced state actors.'' (10)
        \end{itemize}
        \\
    \midrule

    \cite{magic-fsp}
        & AGI Readiness Policy v1.0 \newline (Jul. 2, 2024)
        & \begin{itemize}[itemsep=8pt]
            \item ``\ldots non-expert malicious actor\ldots or an expert\ldots'' (5)
        \end{itemize}
        & \begin{itemize}[itemsep=8pt]
            \item ``\ldots malicious expert actor\ldots'' (4)
            \item ``\ldots a
          talented Computer Science undergrad level malicious actor spending
          3 months and \$1m in compute has a substantial chance of breaking
          critical infrastructure\ldots'' (5)
        \end{itemize} \\
    \midrule

    \cite{meta-fsp}
        & Advanced AI Scaling Framework v2.0 \newline (Apr.\ 8, 2026)
        & \begin{itemize}[itemsep=8pt]
            \item ``\ldots low and moderate skill actors'' (21)
            \item ``Small cells
          of low or moderate skill actors\ldots'' (21) \item ``\ldots
          high-skilled actors'' (21)
          \item ``A well-resourced group of
          high-skilled actors\ldots'' (21)
          \item ``A group with extensive
          resources\ldots'' (21)
        \end{itemize}
        & \begin{itemize}[itemsep=8pt]
            \item ``\ldots would take human expert teams one month or more, on
          average, to find, for significantly less cost and/or time\ldots''
          (19)
        \end{itemize} \\
    \midrule

    \cite{microsft-fsp}
        & Frontier Governance Framework \newline (Feb.\ 8, 2025)
        & \begin{itemize}[itemsep=8pt]
            \item ``\ldots an existing expert (PhD level education in related
          fields)\ldots'' (11)
          \item ``\ldots expert's (PhD level education in related fields) ability \ldots'' (11)
          \item ``\ldots medium-skilled actor's
          (e.g., STEM education) ability\ldots'' (11)
          \item ``\ldots expert's ability \ldots'' (11)
          \item ``\ldots low-skilled actor\ldots'' (11)
        \end{itemize}
        & \begin{itemize}[itemsep=8pt]
            \item ``\ldots ability of a low-skilled actor\ldots'' (12)
          \item ``\ldots low-to-medium skilled actor's ability\ldots'' (12)
          \item ``\ldots low-skilled actor's ability \ldots'' (12)
          \item ``\ldots a well-resourced and expert actor\ldots'' (12)
        \end{itemize} \\
    \midrule

    \cite{naver-fsp}
        & AI Safety Framework \newline (Aug.\ 7, 2024)
        & N/A
        & N/A \\
    \midrule

    \cite{nvidia-fsp}
        & Frontier AI Risk Assessment \newline (Feb.\ 17, 2025)
        & N/A
        & N/A \\
    \midrule

    \cite{openai-preparedness-framework}
        & Preparedness Framework v2.0 \newline (Apr.\ 15, 2025)
        & \begin{itemize}[itemsep=8pt]
            \item ``\ldots `novice' actors (anyone with a basic relevant technical
          background)\ldots'' (5)
          \item ``\ldots non-state actors\ldots''
          (5)
          \item ``\ldots an expert\ldots'' (5)
          \item ``\ldots terrorists\ldots'' (16)
        \end{itemize}
        & \begin{itemize}[itemsep=8pt]
            \item ``\ldots unilateral actors\ldots'' (6)
          \item ``\ldots professional hackers\ldots'' (16)
        \end{itemize} \\
    \midrule

    \cite{xai-fsp}
        & Frontier Artificial Intelligence Framework \newline
          (Dec.\ 31, 2025)
        & N/A
        & N/A \\

\end{longtable}

Some developers sketch relatively detailed profiles -- as in case of \cite{magic-fsp} that specifies the degree of educational attainment, time horizon, and financial capacity -- while others invoke labels such as ``low skilled" and ``well-resourced" without further elaboration on what those tiers entail. Some policies simply employ categorical distinctions (e.g., ``terrorists"), and a meaningful subset name only the harm class and leave the actor entirely implicit. As such, across the industry's current efforts to characterize threat actors, there is variation in both kind and degree that matters for downstream evaluation design. A descriptive persona, however partially, provides a referent against which evaluations can be calibrated; a categorical label or implicit assumption does not. 

This variance is worth highlighting because the same policies that leave their imagined threat actors underspecified also affirm aspirations toward realism. \cite{anthropic-rsp-v3.1} writes that they ``prefer to focus on\ldots what sorts of actors it should address) regarding the risk level from its systems"; \cite{meta-fsp} seeks to ``model competent, incentivized adversaries whose capabilities reflect the specific deployment context" while taking ``into account monetary costs as well as\ldots access to compute, restricted materials, or lab facilities"; \cite{microsft-fsp} specifies that ``[r]esources applied to elicitation should be extrapolated out to those available to actors"; \cite{openai-preparedness-framework} look ``to approximate the high end of expected elicitation by threat actors attempting to misuse the model" and tailor it to ``the level of expected access." 

Realism is an actor-relative property, and the less specified the potential actor, the more the realism standard collapses into a free-floating aspiration rather than a checkable criterion. Linking evaluation design choices back to potential threat actors presupposes that the given characterization is specific enough to support such mapping. For a meaningful portion of the policies surveyed, many important questions remain unanswered. For example, what qualifies an actor to be an expert? What does it mean to be well-resourced? What differentiates medium skill from low skill? And the problem compounds at the mechanistic level of evaluation design: does $pass@100$ or $pass@5$ better reflect the time horizon and financial capacity an actor can sustain? 

Perhaps just as important from Table \ref{tab:threat_actor_characterizations}'s content is what it omits. The developers most prominently represented are ones creating closed-weight models. Major open-weight model developers like Minimax or DeepSeek have largely not participated in public threat actor disclosure. Thus, the players with the greatest post-release exposure are also those with the least developed public language, and, maybe, strategy in turn, for characterizing and guarding against the vast array of potential adversaries.

\chapter{Related Work}

Independent researchers have noted substantial gaps within the nascent risk management practices in the AI industry, including the lack of explicit quantitative risk tolerances and incomplete systemic risk identification \citep{frontier-ai-risk-gap, risk-agi-assessment, iaps-risk-improvement}. Risk identification, as regarded in ISO 31010, encompasses not only detecting what risks exist but also understanding the nature of their sources and the scenarios through which they get realized \citep{iso-risk-management}. This definition situates adversary-focused reasoning as a natural, crucial component of risk identification -- one that current AI evaluation frameworks have begun to incorporate, though not yet systematically. Prior work most relevant to this gap comes from two directions: cybersecurity frameworks for analyzing adversary behavior, and scattered efforts for profiling potential adversaries. 

\section{Analyzing Adversary Behavior}
For more than a decade, the cybersecurity field has approached adversary analysis via structured frameworks organizing observable actor behavior. Lockheed Martin's Cyber Kill Chain decomposes cyberattacks into seven stages: Reconnaissance, Weaponization, Delivery, Exploitation, Installation, Command and Control (C2), and Actions on Objections \citep{cyber-attack-chain-lockheed}. Defense is then framed as the efforts that disrupt an attacker's progression through that chain. The model is fundamentally behavioral, describing what adversaries do and when, rather than characterizing who they are and what they might have. Derived empirically by cybersecurity experts from years of lessons on the job, the Diamond Model of Intrusion Analysis extends the chain by projecting each intrusion event across four features (adversary, infrastructure, capability, and victim), supporting the relationship mapping necessary to effectively structure knowledge about an attack \citep{diamond-model-intrusion}. While the adversary appears as one of its four elements, the Diamond Model functions more as an analytic lens than a substantive account of actor analysis; in other words, it clarifies investigation by organizing perspectives rather than laying out possible actor classes. 

The MITRE ATT\&CK is a knowledge base of adversary tactics drawn from real-world incident data, commonly leveraged to emulate threat actors, plan for red teaming, and assess defensive gaps \citep{mitre-attack}. As such, ATT\&CK is retrospective and observational. MITRE ATLAS \footnote{See https://atlas.mitre.org/.} extends this structure to the AI domain, cataloging attack patterns for AI-enabled systems. The common thread across these efforts is characterizing actors by what they have done, with defense organized around disrupting the observed pathways of action. While this unit of analysis may be conducive for incident response and post-incident attribution, it is less suited to the pre-release context for novel frontier harms, where the need is to reason prospectively about adversary populations that have not yet acted. 

\section{Profiling Potential Adversaries}
Work that addresses the characteristics of potential threat actors themselves is more scattered, and typically embedded ad hoc within analysis aimed at other ends. The RAND Corporation's work on model weight security defines five operational capacity tiers to distinguish between adversary groups when assessing the risk of model weight ex-filtration: (1) amateur attempts, (2) professional opportunistic efforts, (3) cybercrime syndicates and insider threats, (4) standard operations by leading cyber-capable institutions, (5) top-priority operations by the top cyber-capable institutions \citep{rand-secure-model-weights}. In describing each of these groups, the report considers operation headcount, domain knowledge, infrastructure, monetary resources, and time horizon. For example, a professional opportunistic effort describes cases where a single individual proficient in information security has a few weeks, a budget up to \$10,000, and has personal cyber equipment.

The Structured Threat Information Expression (STIX) 2.1 threat actor vocabularies \footnote{See sections 10.23 through 10.25: https://docs.oasis-open.org/cti/stix/v2.1/cs02/stix-v2.1-cs02.html.} provide a schema for representing threat actors via type, role, and sophistication. The threat actor type spans twelve terms, such as criminal, hacker, and terrorist. The role distinguishes seven functions: agent, director, independent, infrastructure-architect, infrastructure-operator, malware-author, and sponsor. The sophistication gradient features seven levels: none, minimal, intermediate, advanced, expert, innovator, strategic. Despite this object-level attention to detail, these fields are primarily designed to be reserved as metadata labels for threat intelligence sharing rather than as a graded analytical scale for evaluation calibration. The Centre for Long-Term Resilience's work on assessing AI's near-term biological misuse impacts deconstructs threat actors along several dimensions, separating individuals from groups, and scaling each wih respect to their capabilities in AI research, foundational model training, model customization, and biological research \citep{cltr-bio-misuse}. These efforts are valuable but remain domain-specific.

Taken together, the existing literature occupies two positions. Frameworks for analyzing adversary behavior are well-developed but restrictively retrospective. Work that profiles potential adversaries exists but remains fragmented -- distributed across policy documents, biosecurity scholarship, and intelligence taxonomies not designed to interoperate. To the best of our knowledge, there has not been an undertaking to standardize analyzing the adversary landscape in the context of both CBRN and offensive cyber AI misuse risk. We address this by distilling meaningful concerns about threat actors into six attributes organized into a generalizable taxonomy. In doing so, our framework not only provides a common language for different model developers to articulate the threat actors they are concerned about, but also a tool for linking evaluation design to those very concerns -- so that results are ultimately more faithful to the target risk profile.

\chapter{Standardizing Threat Actor Characterization}
\label{chp:4}

The preceding chapters have documented two related gaps: evaluation practices that invoke assumptions about threat actors without making them explicit, and a research landscape that lacks common grounding concepts needed to compare experiments conducted by different institutions. At root, both are standardization problems, as capability evaluations do not produce context-free measurements, and treating them as such obscures what their results actually mean \footnote{The Environmental, Social, and Governance (ESG) domain illustrates the consequences when multiple bodies, without a shared framework, independently select attributes to characterize a complex phenomenon. \cite{esg-confusion} find that across six major rating agencies, ESG ratings differ substantially (correlations range from 0.38 to 0.71), leading to ``aggregate confusion." Each rater chose different sets of attributes for assessment, and this kind of scope difference contributes to 38\% of the rating divergence. The agencies' rating methodologies were sufficiently incompatible that the authors had to construct a common taxonomy retroactively to compare them.}. A growing body of practitioner experience and policy analysis has underscored that results are shaped by the implicit assumptions embedded in evaluation design that are often overlooked -- which tools are available in a given lab environment, what are the participants' AI literacy levels, how much time is allotted for a task \citep{measurement-to-meaning, methodological-challenges, ai-rct-paper}.

Recent work from the RAND Corporation on AI-biology evaluations stresses that even the smallest of differences in participant expertise, available resources, and contextual constraints can have large and unexpected impact on observed performance -- effects that compound when studies are compared with without shared, understood reference points to pin down those assumptions \citep{rand-bio-measure}. Clear and explicit documentation can help mitigate risks of  misinterpretation and overgeneralization; the absence of such pre-commitment has manifested as reporting bias in medicine and economics, fields that have long employed RCT-based methodology \citep{minimizing-rct-risks}.

Joint analyses produced through the U.S.-China Track II Dialogue on AI and National Security, from the Brookings Institution's Foreign Policy Program and Tsinghua University's Center for International Security and Strategy, perhaps help identify the right unit of analysis for this standardization effort \citep{brookings-non-state-actor}. They argue that misuse risk is not a single-dimensional property of a model's capabilities, but emerges from the interaction of many dimensions, such as an actor's baseline skills, usable infrastructure, and surrounding governance regimes. Proper attention must therefore be given to this wider contextual risk, not solely capability risk. The implication is that the actor should be the variable that evaluation design must be anchored to, not the model nor the harm category alone. 

A natural challenge to standardization is that different organizations adopt different threat models, making a shared framework seem overly constraining. However, this objection underestimates a subtle but systemic problem: when capability evaluations are built on divergent adversary assumptions, their results become incomparable in ways that are not always visible. Two labs may appear to be measuring the same risk while actually reasoning from different threat premises. This matters especially for marginal risk assessment, where without some shared adversary baseline, the comparison lacks a stable reference point \citep{estimating-worst-case-risk, opus-46-model-card}. Effectively, each attempt is measuring a delta against a different zero. Crucially, standardizing adversary analysis does not expect nor require agreements about threat models; rather, it surfaces disagreements explicitly, creating the common reference points necessary for risk analyses to be meaningfully compared, aggregated, and acted upon across the field. Two claims follow: evaluation results that do not specify the actor they are calibrated against cannot be reliably interpreted, and without a common framework for specifying that actor, they cannot be compared across experiments either. The rest of this paper describes our attempt to reconcile these issues.

\section{Taxonomy Design Goals}

The taxonomy is designed to be generalizable (yet still adequately fine-grain and comprehensive), systematic, and operationalizable:
\begin{itemize}
    \item \textbf{Generalizability} ensures that the framework applies across the full range of actors within CBRN and offensive cyber AI misuse risk, without requiring much domain-specific adaptation. Attributes are derived from existing concerns that recur across CBRN and cyber threat analysis. In this way, we preserve cross-domain stability, making the resulting taxonomy useful as a common language across frontier AI misuse evaluations.
    \item \textbf{Systemacity} requires that the taxonomy prompts stakeholders to consider every relevant dimension of an actor, not only the ones that feel most salient to a given evaluation context. A matrix format serves this purpose: by presenting all attributes as explicit fields to be filled in, the structure renders omissions visible rather than implicit. More formally, this could be viewed as a structural analogue to the ``pre-analysis plans" utilized as standard practice in many medical and economic journals to combat reporting bias -- the framework not only describes what to consider, but also compels a discipline of considering it \citep{minimizing-rct-risks}. 
    \item \textbf{Operationalizability} requires that the framework support concrete evaluation design choices rather than stalling at high-level descriptions. Each attribute is specified in tiers with observable criteria, so that a given actor profile translates into checkable assumptions (e.g., what elicitation methods are realistic, or what time constraints should be imposed on a task). Additionally, the attribute-based modular structure can be particularly fruitful. As attributes can be varied independently while others are held constant, engaging with the taxonomy can surface and accommodate nonobvious edge cases that holistic, top-down characterizations could miss. For instance, a small cell of homegrown extremists might score low on financial capacity and operational infrastructure but high on domain knowledge and task time horizon \citep{biothreat-benchmark}. 
\end{itemize}

\section{Attribute Derivation}

To characterize threat actors meaningfully for AI risk evaluations, we distinguish six attributes: technical sophistication, prior domain knowledge, organizational capacity, operational infrastructure, financial capacity, and time horizon. These attributes were not constructed via first principles but distilled through a systematic review of existing CBRN and cyber threat analysis. Namely, we apply \cite{taxonomy-dev}'s method for our taxonomy development, which involves determining a meta-characteristic\footnote{Defined as ``the most comprehensive characteristic that will serve as the basis for the choice of characteristics in the taxonomy" \citep{taxonomy-dev}.} and ending conditions, and iterating between conceptual or empirical derivation approaches until those conditions are satisfied. For the purpose of classifying threat actors in frontier AI misuse, our meta-characteristic is defined as the capacities that would be helpful to adversaries to use and cause harm with AI systems. We follow the eight objective and five subjective ending conditions outlined in \cite{taxonomy-dev}, which we assess in Section \ref{sec:5-assess-quality} after the full taxonomy and derivation details are presented.

In determining the high-level attributes, four bodies of work support our conceptual-based approach: RAND's operational capacity definitions from their paper on securing model weights \citep{rand-secure-model-weights}, the biological weapon risk chain highlighted in a separate RAND report on measuring biological capabilities of AI agents \citep{rand-bio-measure}, a US government-adjacent analysis on deterring weapons of mass destruction (WMD) \citep{us-cbrn-terrorism-2001}, and Lockheed Martin's Cyber Kill Chain \citep{cyber-attack-chain-lockheed}. While the attributes are not always named consistently across these sources, their underlying themes recur. 

Technical sophistication emerges across all sources, each of which attends in some form to how well a threat actor understands and can deploy the underlying tools relevant to their objective. Prior domain knowledge comes from the biosecurity and WMD deterrence literature, which assume some baseline understanding of the relevant subject matter as a precondition for operational capability. Organizational capacity is foregrounded in RAND's operational capacity tiers, which treat group structure and headcount as important variables in assessing adversary capability. Operational infrastructure appears across both the Cyber Killer Chain and the biological weapon risk chain, where access to certain equipment and materials are concerns at various stages. Both financial capacity  and time horizon are also addressed in RAND's operational capacity definitions, where each is rationalized and divided among tiers. 


\chapter{Decomposing Threat Actors: An Attribute Taxonomy}
\label{chp:5}
 
\begin{figure}[!htbp]
  \centering
  \includegraphics[width=0.8\linewidth]{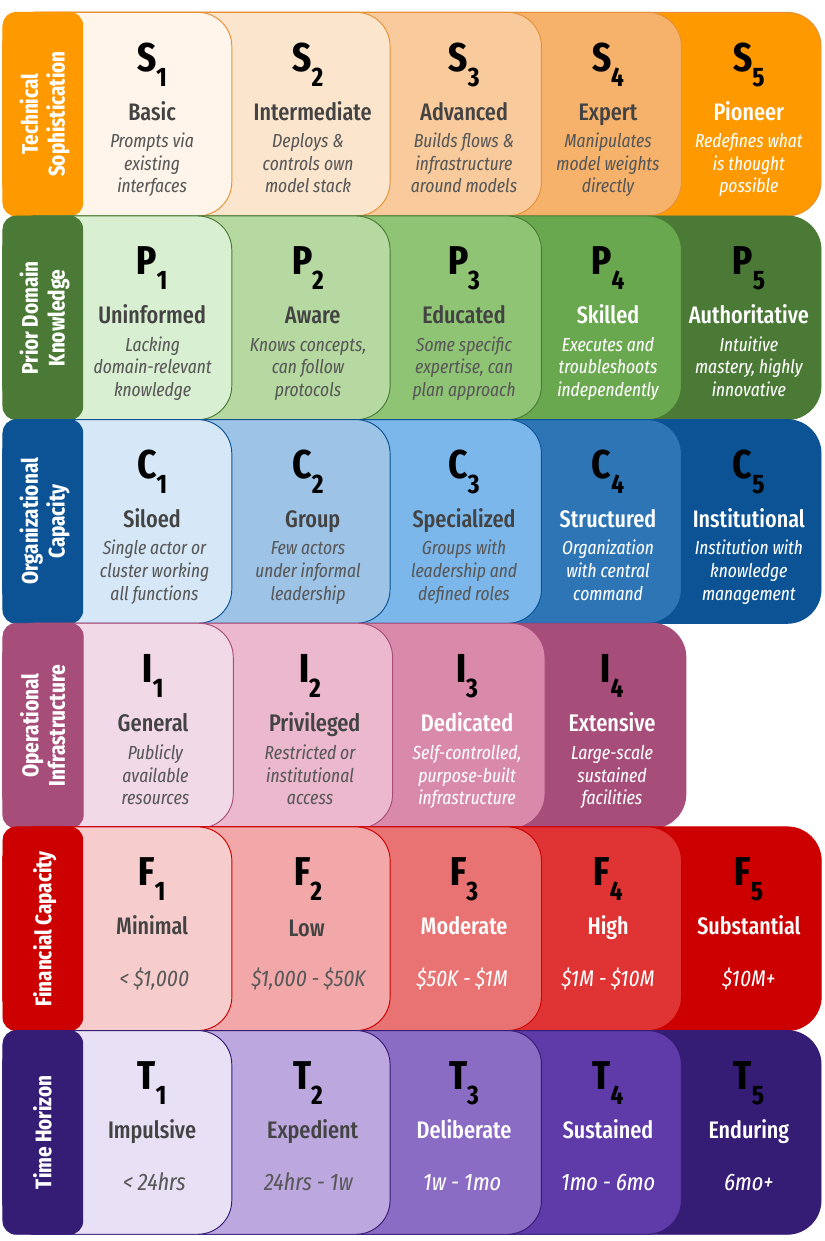}
  \caption[Threat Actor Characterization Matrix]{A six-attribute adversary characterization matrix for frontier AI misuse risk. Each attribute is divided into tiers progressing from least to most capable or resourced. }
  \label{fig:full_matrix}
\end{figure}

Figure \ref{fig:full_matrix} presents the full threat actor characterization matrix. Each row corresponds to one of six attributes, and each column corresponds to a tier within a single attribute, progressing from least to most capable or resourced from left to right. A complete threat actor profile is a filled vector across all six rows: a specific combination of tiers that describes the structural properties of a hypothetical actor without referencing any particular identity or affiliation. It is the combination of all attributes that matters -- no single tier on any attribute should determine an overall risk profile. The matrix is designed to make these combinations explicit and comparable across evaluators. In the following sections, we explain each attribute in depth, detailing its relevance to risk assessment and the basis for its tier gradient. Using \cite{taxonomy-dev}'s terminology, conceptual approaches were the primary mode for all attributes except financial capacity and time horizon, where empirical approaches were used.

\section{Technical Sophistication}

\textit{The abilities an actor has for interacting with, hosting, extending, or modifying an AI model.}

\begin{figure}[!htbp]
    \centering
    \includegraphics[width=0.8\linewidth]{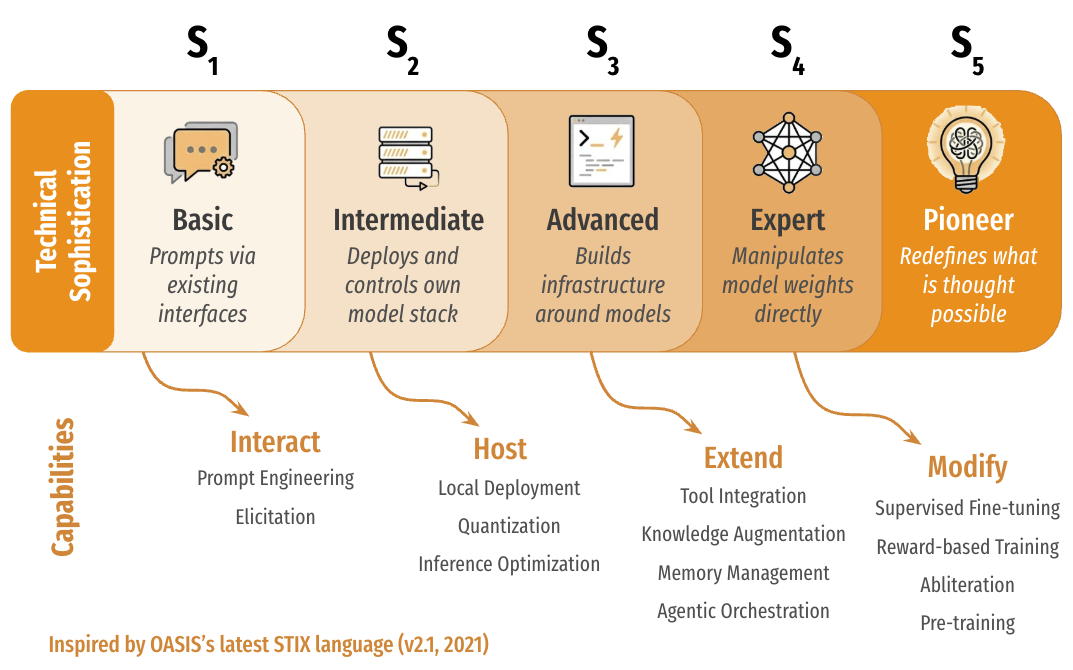}
    \caption[Technical Sophistication Attribute Explainer]{Technical sophistication tiers, with associated capabilities at each level. Tier labels reflect an actor's highest available mode of AI interaction, progressing from prompt-based engagement ($S_1$) through local deployment and infrastructure-building to direct weight manipulation and novel capability development ($S_4$–$S_5$). Capability subcategories are illustrative rather than exhaustive. Inspired by the OASIS Open STIX 2.1 sophistication vocabulary.}
    \label{fig:tech_sophistication}
\end{figure}

\textbf{Why It Matters}. Technical sophistication can help approximate both the feasibility and ceiling of misuse. Many high-impact failure modes are not accessible through naïve interaction with a model; they require the ability to systematically probe, adapt, and chain capabilities over time. More technically sophisticated actors can bypass safeguards, integrate models into larger operational pipelines, and extract marginal gains from partial capabilities that less technically capable users would not realize \citep{gdm-ai-cyber-framework, best-bio-practices}. As a result, considering technical sophistication helps clarify which risks are likely to manifest under certain conditions versus those contingent on more advanced actors. 

\textbf{Reasoning Behind Tiers}. The tier structure draws on the technical sophistication model defined in the OASIS Open STIX 2.1 standard. STIX is a standard language for cyber threat intelligence sharing across government and industry, developed under OASIS Open, a prominent non-profit standards body for international policy \footnote{For more information about Oasis Open, see https://www.oasis-open.org/org/.} Its technical sophistication model organizes threat actor capability around a sequence of escalating checkpoints: from using existing tools without understanding them, through setting up custom workflows, to discovering and exploiting unknowns. That progression is preserved here, reframed  around AI-specific technical capability and validated against practitioner curricula and academic courses (DeepLearning.AI \footnote{http://DeepLearning.AI}, DeepWiki \footnote{https://deepwiki.com/mlabonne/llm-course}, Stanford CS224n\footnote{https://web.stanford.edu/class/cs224n/}). 

The resulting tiers (see Figure \ref{fig:tech_sophistication}) are roughly cumulative and function as a heuristic rather than a rigid typology. A \textit{Basic} actor can direct a model with intention through existing interfaces, relying on prompt engineering and elicitation. An \textit{Intermediate} actor additionally controls their own deployment infrastructure (e.g., running models locally, managing quantization, and optimizing inference) rather than depending on external assistance. An \textit{Advanced }actor can build novel pipelines and workflows around a model, integrating external tools and APIs, augmenting model knowledge, managing persistent context, and orchestrating multi-step autonomous workflows. An \textit{Expert} actor can manipulate the model weights directly, through supervised fine-tuning, reward-based training, abliteration, or pre-training. A \textit{Pioneer} actor operates at the frontier of what the field considers achievable, unlocking capabilities that are not merely novel applications of existing methods but breakthroughs that expand the boundary of the possible.

\section{Prior Domain Knowledge}

\textit{The extent to which an actor already possesses expertise in a harm-relevant domain (i.e. biological, chemical, radiological, nuclear, or offensive cyber) independent of what AI can provide.}

\begin{figure}[!htbp]
    \centering
    \includegraphics[width=0.8\linewidth]{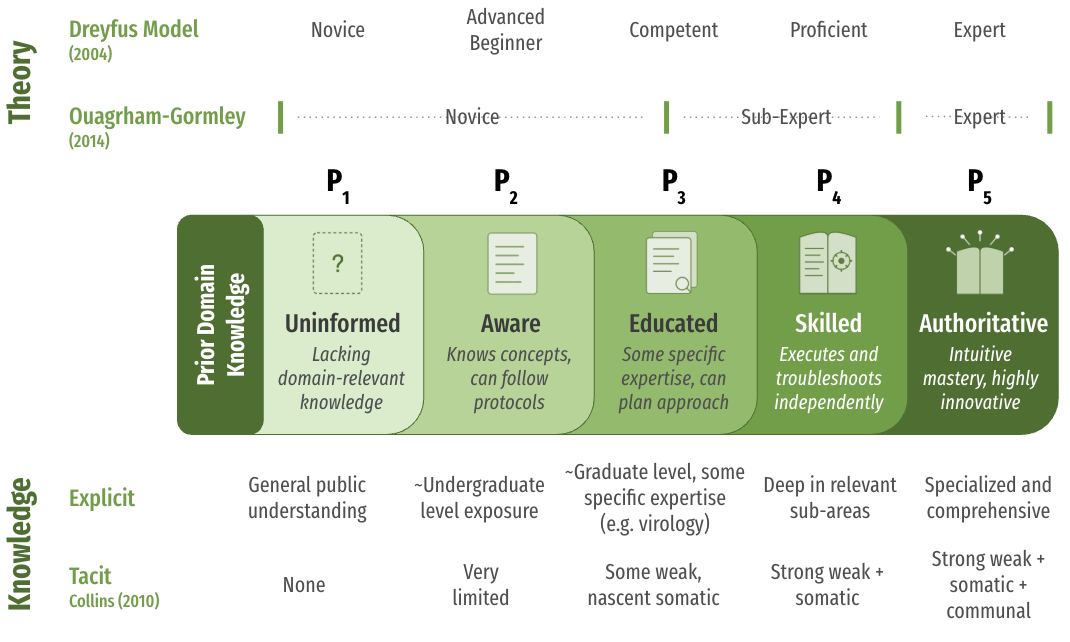}
    \caption[Prior Domain Knowledge Attribute Explainer]{Prior domain knowledge tiers, mapped against two theoretical frameworks. Above the tiers, the \cite{dreyfus-skill} Model of skill acquisition and \cite{ouagrham-bioweapons}'s novice–sub-expert–expert typology provide theoretical grounding for tier boundaries. Below, each tier is characterized along two knowledge dimensions: explicit knowledge, reflecting codified understanding of domain concepts and techniques, and tacit knowledge, following \cite{collins-tacit-knowledge}'s taxonomy of weak, somatic, and communal forms. Tier labels are domain-agnostic and apply across CBRN and cyber harm categories.}
    \label{fig:prior_domain_knowledge}
\end{figure}

\textbf{Why It Matters}. Prior domain knowledge determines how much of an actor's capability gap AI must bridge, and therefore how decisively a given model release changes what they can achieve. An actor who already holds deep expertise in a harm-relevant domain requires only marginal assistance from an AI system to execute complex operations. An actor who lacks such expertise may depend on AI assistance to perform tasks that experts would find routine. The actor's pre-existing competency level also determines the function a model would most effectively serve: collaborator, force multiplier, or knowledge gap filler. Needing to clarify the level and domain of the human's expertise is a commonly recurring lesson learned by evaluators working in this space \citep{rand-bio-measure}.\footnote{See their Section 4.1.4, ``Specifying Level and Domain of Human Expertise."}

\textbf{Reasoning Behind Tiers}. The tier structure (see Figure \ref{fig:prior_domain_knowledge}) is anchored in two bodies of work: the \cite{dreyfus-skill} model of skill acquisition, which describes a trajectory from novice through advanced beginner, competent, proficient, and expert; and \cite{collins-tacit-knowledge}'s typology of tacit knowledge, which distinguishes weak, somatic, and communal tacit knowledge as qualitatively distinct forms of expertise that cannot be reduced to explicit, transmittable content. \cite{revill-jefferson} applies Collins' framework to biological weapons, and \cite{ouagrham-bioweapons} reinforces it, identifying tacit knowledge as a binding constraint on weapons development programs. 

\cite{ouagrham-bioweapons}'s analysis distinguishes three actor categories: novices, who have a generic educational background but no specific or practical expertise; sub-experts, who possess advanced theoretical knowledge and specific disciplinary expertise but lack practical weapons development experience; and experts, who combine advanced theoretical grounding with practical expertise. This is complemented by the Weapons of Mass Destruction Proxy benchmark (WDMP)'s approach to testing for hazardous capabilities; their framework delineates three nested, increasingly hazardous levels of knowledge: general domain knowledge, expert-level disciplinary knowledge, and weapons-specific knowledge \citep{wmdp-benchmark}.  Actors classified as an expert under Ouagrham-Gormley's typology have deep penetration of the most dangerous level of knowledge.

\section{Organizational Capacity}

\textit{The structure of the human network an actor operates within for coordination, specialization, and sustainment.}

\begin{figure}[!htbp]
    \centering
    \includegraphics[width=0.8\linewidth]{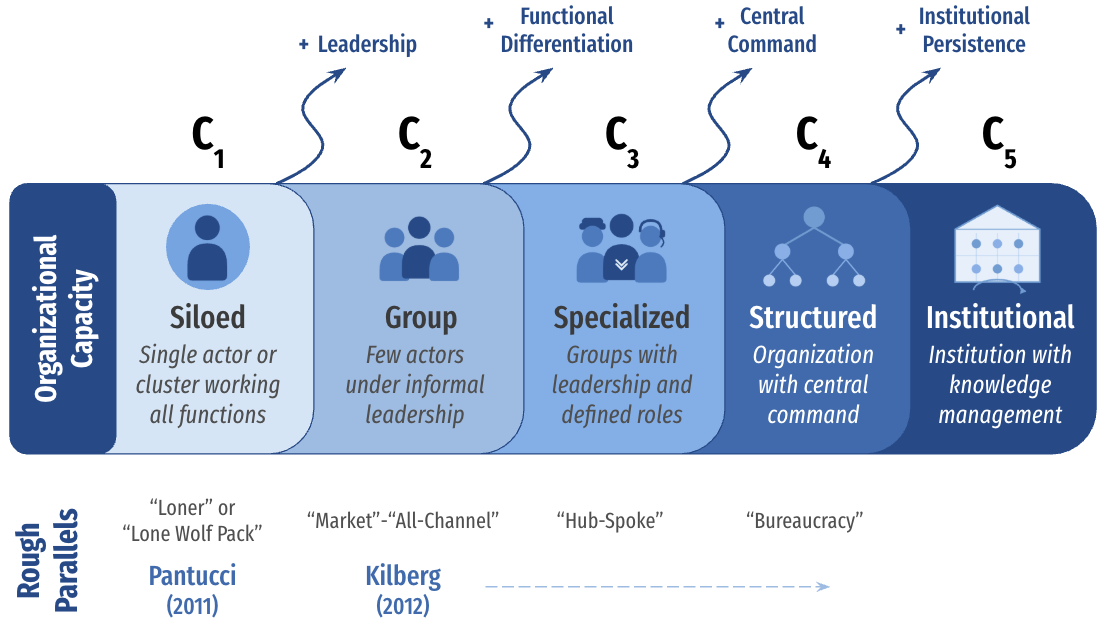}
    \caption[Organizational Capacity Attribute Explainer]{Organizational capacity tiers, each adding exactly one structural property to the previous. Arrows indicate the defining addition at each transition: leadership ($C_1$ $\rightarrow$ $C_2$), functional differentiation ($C_2$ $\rightarrow$ $C_3$), central command ($C_3$ $\rightarrow$ $C_4$), and institutional persistence ($C_4$ $\rightarrow$ $C_5$). Below the tiers, rough parallels are drawn to established typologies in the terrorism studies literature — \cite{pantucci-terrorism}'s loner and lone wolf pack for $C_1$, and \cite{kilberg-terrorism}'s market, all-channel, hub-spoke, and bureaucracy network forms for $C_2$ through $C_4$. $C_5$ has no direct parallel in either typology, reflecting the upper bound of organizational complexity beyond what either framework was designed to capture.}
    \label{fig:organizational_capacity}
\end{figure}

\textbf{Why It Matters}. The structure of the human network an actor operates within directly conditions what they can attempt and sustain. \cite{lethality-terrorism} establish that both organizational size and connectedness are linked to lethality and operational success. A larger and more differentiated organization can specialize labor, sustain operations across failures, accumulate and transmit institutional knowledge, and run parallel workstreams. Such dynamics are unavailable to a single actor or small cluster, regardless of individual technical sophistication. Organizational capacity therefore sets a ceiling on operational ambition that technical factors alone may miss.

\textbf{Reasoning Behind Tiers}. The tier structure is grounded in \cite{kilberg-terrorism}'s model of terrorist organization, which defines four structures along three dimensions: leadership, functional differentiation, and central command. Kilberg's typology was applied to 254 cases drawn from the Global Terrorism Database (1970–2007), giving it empirical grounding across a wide range of real-world organizational forms. The resulting structures (market, all-channel, hub-spoke, and bureaucracy) map onto observable variation in how groups divide labor, exercise authority, and persist across time. These four structures are combined with \citet{pantucci-terrorism}'s decomposition of the solo-actor dimension, distinguishing loners, lone wolves, lone wolf packs, and lone attackers, and the two most relevant spectrums from \citet{borum-lone-offender}: loneness (degree of direct assistance from others) and direction (extent of independence and autonomy in decision-making). Together, they generate the five tiers presented in Figure \ref{fig:organizational_capacity}.

A \textit{Siloed} actor is an individual or small cluster of two to four people performing all functions, with scope limited to what that individual or cluster can hold. A \textit{Group} consists of roughly four to twelve actors operating under informal leadership, with no functional differentiation, no centralized command, and possibly limited to a single operation. A \textit{Specialized} structure involves twelve to thirty actors with defined roles and a few coordinating hub individuals, approximating \cite{kilberg-terrorism}'s hub-spoke model. A \textit{Structured} organization of thirty to one hundred actors adds central command and formal hierarchy, enabling sustained campaigns, recovery from failure, and the beginnings of long-term research and development capacity. An \textit{Institutional} actor, with one hundred or more members, further adds formal knowledge management, institutional memory decoupled from specific individuals, and the capacity to run multiple workstreams in parallel -- the organizational prerequisites for sustained frontier-level R$\&$D. 

\section{Operational Infrastructure}

\textit{The existing assets (physical and digital) an actor can leverage.}

\begin{figure}[!htbp]
    \centering
    \includegraphics[width=0.8\linewidth]{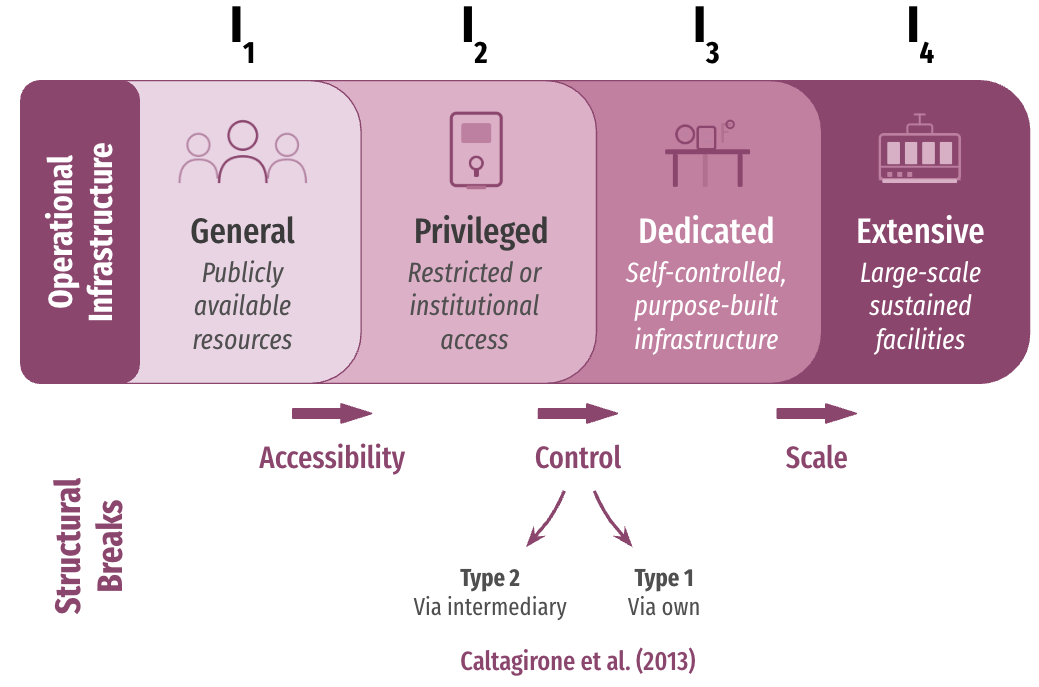}
    \caption[Operational Infrastructure Attribute Explainer]{Operational infrastructure tiers, organized by three structural breaks: accessibility ($I_1$ $\rightarrow$ $I_2$), control ($I_2$ $\rightarrow$ $I_3$), and scale ($I_3$ $\rightarrow$ $I_4$). The control transition is further distinguished by the nature of access (Type 2 infrastructure, accessed via an intermediary, versus Type 1 infrastructure, owned and controlled directly by the actor) following the typology introduced in the Diamond Model of Intrusion Analysis \citep{diamond-model-intrusion}. Unlike other attributes, operational infrastructure incidentally uses four tiers rather than five.}
    \label{fig:operational_infrastructure}
\end{figure}

\textbf{Why It Matters}. Operational infrastructure captures the existing physical and digital assets an actor can leverage, independent of what they could acquire with additional resources. These assets function as enablers that determine which steps of an attack chain are immediately actionable versus which require additional acquisition, a distinction that directly shapes the marginal contribution of AI assistance. Needing to define the resource assumptions is a commonly recurring lesson learned by evaluators working in this space \citep{rand-bio-measure}.\footnote{See their Section 4.1.5, ``Defining Resource Assumptions."}

\textbf{Reasoning Behind Tiers}. As shown in Figure \ref{fig:operational_infrastructure}, the tiers are organized not to catalog every possible type of infrastructure but to identify structural breaks marking capability ceilings imposed by what is physically or technically available to a given actor. The progression is one of layered constraints, each imposing a qualitatively different ceiling: accessibility, then control, then scale.

The first and most significant gate is accessibility. A \textit{General} actor is limited to publicly available resources: consumer devices, public internet, open-source software, freely downloadable datasets, and commercially available compute. The transition to \textit{Privileged} represents the first major threshold: the ability to access specialized or restricted resources not available to the general public, whether through an institutional role, insider position, or other means. University and corporate laboratories, administrative privileges on IT systems, regulated materials, proprietary codebases, and institutional compute access all belong to this tier.

Once the accessibility threshold is crossed, the binding constraint becomes control. A \textit{Dedicated} actor does not merely access restricted resources through a third party that retains the authority to revoke that access; they own and operate self-controlled, purpose-built infrastructure, such as private laboratories, custom software toolchains, dedicated compute clusters, air-gapped systems, and established procurement channels. This distinction between privileged access and full control informs the boundary between the \textit{Privileged} and \textit{Dedicated} tiers and draws directly on the Diamond Model of Intrusion Analysis from cybersecurity, which distinguishes Type 1 infrastructure (fully owned and controlled by the adversary) from Type 2 infrastructure (used by the adversary but owned by an intermediary) \citep{diamond-model-intrusion}. 

Above the control threshold, scale becomes the determinant. An \textit{Extensive} actor operates large-scale, sustained infrastructure with purpose-specific facilities (e.g., BSL-3 or BSL-4 biological containment, cyber operation centers, frontier-scale compute, sustained supply chains, and manufacturing plants). At this tier, the actor has access to nearly every resource that infrastructure can provide at the individual scale and above, enabling sustained, complex, and parallel operations that lower-tier actors simply cannot run.

\section{Financial Capacity}

\textit{The capital available to an actor for acquiring resources and services relevant to carrying out objectives.}

\begin{figure}[!htbp]
    \centering
    \includegraphics[width=0.8\linewidth]{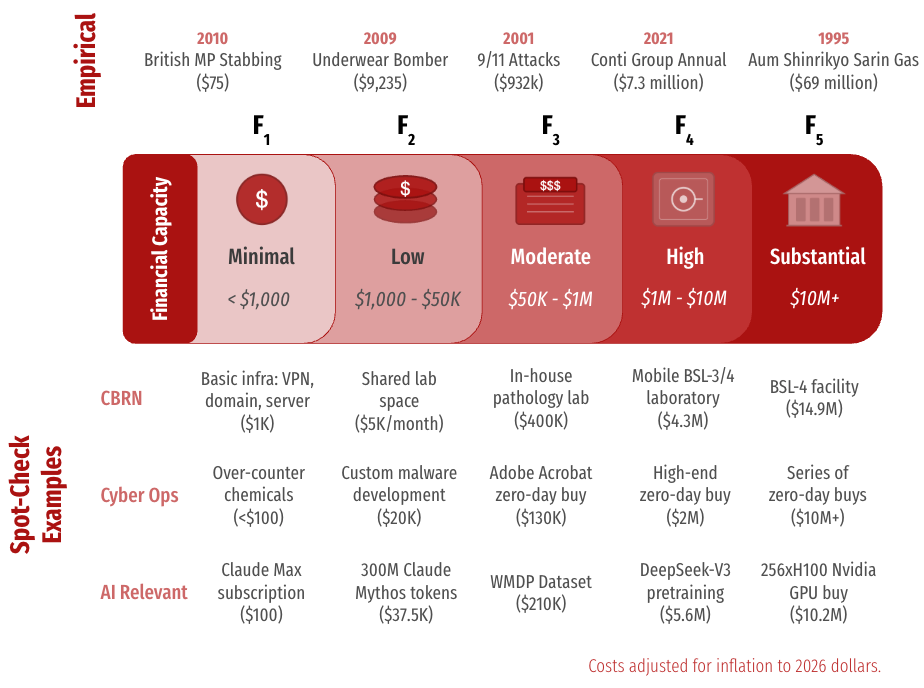}
    \caption[Financial Capacity Attribute Explainer]{Financial capacity tiers, with empirical anchoring and spot-check examples. Above the tiers, real-world operational costs from terrorism and cybercrime incidents ground each tier boundary, ranging from the 2010 British MP stabbing (\$75) at $F_1$ to the 1995 Aum Shinrikyo sarin attack (\$69 million) at $F_5$. Below, illustrative costs are provided across three categories (CBRN acquisition, cyber operations, and AI-relevant expenditures) to validate tier boundaries against current market prices. All costs are adjusted for inflation to 2026 dollars.}
    \label{fig:financial_capacity}
\end{figure}

\textbf{Why It Matters}. Financial capacity is analytically distinct from operational infrastructure. Infrastructure captures what the actor currently controls; financial capacity captures what they can acquire. It models the latitude an actor has to expand their capabilities through purchases (whether of compute, laboratory access, exploits, personnel, or external services) and therefore governs the ceiling of their operational reach beyond their existing baseline. Recent work from the UK AI Security Institute on inference scaling for cyber tasks finds that higher token budgets correlate with higher success rates, highlighting that previous evaluations, by operating under frugally conservative token budgets, have systematically assumed underestimates of model-assisted adversarial capability \citep{UK-aisi-more-inference-scaling}. As such, financial capacity shapes what an AI-assisted actor can accomplish in ways that evaluation methodology must explicitly account for.

\textbf{Reasoning Behind Tiers}. The tier boundaries are derived empirically by identifying natural clusters in data on how much threat actors have actually spent on operations, then spot-checked against the costs of AI compute, traditional cyber capabilities, and CBRN-relevant laboratory equipment at each level. The primary empirical source is \cite{terrorist-financing}'s analysis of forty terrorist plots in Western Europe between 1994 and 2013, which finds that nearly all fell in what this framework designates as the \textit{Low} tier or below (see Figure \ref{fig:financial_capacity}). This empirical floor is complemented by data covering the September 11 attacks, and by financial records on Aum Shinrikyo's sarin synthesis program \citep{monograph-us-terrorist-finances, aum-analysis}. Additional anchors come from our own estimates based on law enforcement disclosures and counterterrorism financing reports for individual incidents found online.

The \textit{Minimal} tier, below one thousand dollars, is calibrated against cases such as the 2010 stabbing of British MP Stephen Timms (approximately at maximum seventy-five dollars)\footnote{For more information about the attack, see https://en.wikipedia.org/wiki/Stabbing\_of\_Stephen\_Timms.}, crude improvised CBRN precursor materials available over the counter for under one hundred dollars \footnote{See https://www.chemworld.com/Default.asp for an example.}, and consumer API access to frontier AI models \footnote{For example, Claude Pro and Max subscription tiers are priced at \$20 and \$100 monthly, respectively as of April 21, 2026. See https://claude.com/pricing.}. The \textit{Low} tier, spanning one thousand to fifty thousand dollars, encompasses incidents such as the 2006 German train bombing attempt and the 2009 Underwear Bomber \citep{terrorist-financing}, as well as shared commercial laboratory space \citep{rent-lab-small}, basic malware packages, and substantial API usage at current pricing \footnote{For example, Multi-modal Kimi K2.5 output pricing (per million tokens) is \$3 as of April 21, 2026. See https://platform.kimi.ai/docs/pricing/chat-k25.}. 

The \textit{Moderate} tier, from fifty thousand to one million dollars, covers the September 11 attacks at roughly nine hundred thousand dollars (inflation-adjusted to 2026), the construction of in-house pathology laboratories \citep{in-house-pathology-lab}, and the acquisition of a dedicated GPU cluster for open-weight model fine-tuning \footnote{Purchasing an NVIDIA H100 (Hopper architecture, 80 GB HBM3 is roughly \$40,000: https://intuitionlabs.ai/articles/nvidia-ai-gpu-pricing-guide.}. The \textit{High} tier, from one to ten million dollars, corresponds to operations like Conti Group ransomware campaigns \citep{conti-group-analysis}, mobile BSL-3/4 laboratory construction \footnote{See https://qualia-bio.com/blog/mobile-bsl-3-bsl-4-module-labs-a-cost-effective-alternative/ for estimates.}, and large-scale malicious fine-tuning of open-source models \citep{estimating-worst-case-risk}. The \textit{Substantial} tier, above ten million dollars, reflects nation-state and near-state-level capabilities: Aum Shinrikyo's sarin synthesis program at roughly sixty-nine million dollars (inflation-adjusted to 2026) \citep{aum-analysis}, full BSL-4 facility construction \citep{full-bsl-4-facility}, and frontier-model pre-training \citep{frontier-model-training-costs}. Since data is relatively sparse at the upper end of the scale, the \textit{Substantial} tier in particular should be treated as a preliminary attempt, subject to revision as more evidence becomes available. The limited empirical record for high-end operations reflects both the rarity of such activities and the difficulty of reconstructing their financing from available sources.

\section{Time Horizon}

\textit{The duration over which an actor pursues a harmful objective, including preparatory work.}

\begin{figure}[!htbp]
    \centering
    \includegraphics[width=0.8\linewidth]{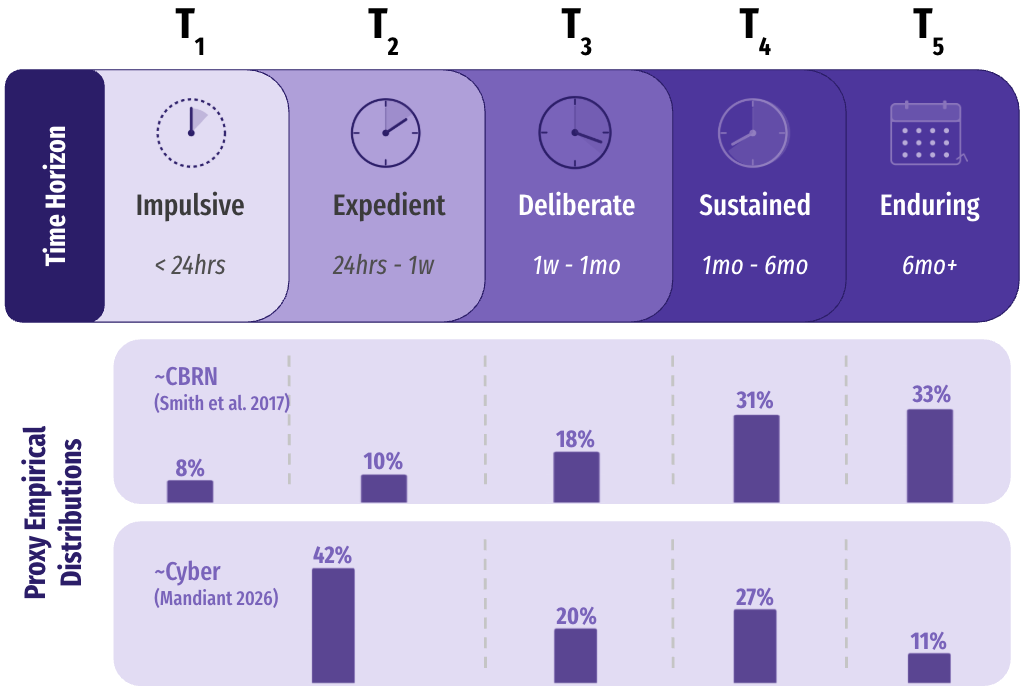}
    \caption[Time Horizon Attribute Explainer]{Time horizon tiers, with proxy empirical distributions across two harm domains. Tier boundaries are defined by planning and execution duration, ranging from impulsive operations under 24 hours ($T_1$) to enduring campaigns of six months or more ($T_5$). Below, bar charts show the approximate distribution of real-world incidents across tiers for CBRN-proximate attacks, drawn from Smith et al.'s (2017) American Terrorism Study quartile data, and for cyber intrusions, drawn from Mandiant's M-Trends 2026 dwell time data. These distributions are treated as proxies rather than direct measurements, as neither dataset was designed specifically to capture AI-enabled misuse planning horizons.}
    \label{fig:time_horizon}
\end{figure}

\textbf{Why It Matters}. Time horizon captures the duration over which an actor pursues a harmful objective, including all preparatory work. A scaling dynamic is at work: extended time allows for iterative refinement, the accumulation of domain knowledge, and the kind of persistent engagement needed to navigate complex multi-step tasks. Longer time horizons also expand the range of operational options available; for example, a patient actor can wait for favorable conditions, conduct reconnaissance, and recover (and learn) from failed attempts in ways that a heavily time-constrained actor cannot.

\textbf{Reasoning Behind Tiers}. Two empirical sources anchor the tier structure (see Figure \ref{fig:time_horizon}). For physical and CBRN risks, \cite{smith-terrorism-plan-time} provide temporal data on terrorist planning cycles drawn from the American Terrorism Study (1980–2016). Of the 1,360 cases in the dataset, 272 contained usable temporal information, and the distribution reveals that planning cycles span a wide range: roughly five percent of cases involved less than one day of planning; one in five planned within ten days; the interquartile range runs from twenty days to approximately ninety-five days; and almost all cases (98\%) concluded within three years. Breaking this distribution into quartiles (zero to 20 days, 21 to 95 days, 96 to 285 days, and 286+ days) provides the empirical scaffold for the CBRN-relevant tier boundaries.

For cyber risks, \cite{madiant-m-trends-2026}'s annual M-Trends report provides a complementary empirical anchor through the concept of dwell time: the number of days an attacker is present in a compromised environment before detection. Based on analysis of approximately five hundred thousand hours of frontline incident investigations conducted by the Mandiant team in collaboration with Google's Threat Intelligence Group, the global median dwell time in 2025 was fourteen days. Specifically, the distribution shows that about 42\% of intrusions had dwell times within one week, 20\% between 8 and 30 days, 27\% between 31 days and 6 months, 6\% greater than 6 months but less than one year, and 6\% greater than one year.

We acknowledge that these two sources do not measure the same phase of an operation: \cite{smith-terrorism-plan-time}'s data tracks pre-incident preparatory activities, while the \cite{madiant-m-trends-2026} data captures active intrusion dwell time. The reason for this may be partly inherent to the mode of attack: digital forensics typically begins at the point of initial access, since pre-intrusion phases (reconnaissance, weaponization, infrastructure development) may leave minimal to no traces that can be attributed before an incident is confirmed. Physical terrorism, by contrast, eventually produces court records, surveillance footage, financial transactions, and witness testimony that allow researchers to reconstruct preparation timelines after the fact. Despite this asymmetry, we use both sources as relative proxies to calibrate the tier boundaries, and do not regard them as precise measurements of equivalent phenomena. 

Additionally, given that current AI tools may meaningfully compress attack timelines by functioning as force multipliers \citep{mythos-model-card},\footnote{See their Section 2.2.5.1 on expert red-teaming for biological risks, where experts assessed the model as a useful force multiplier for speed and breadth of research.}, the tier structure includes granular initial tiers at the sub-day and one-to-seven-day level to capture the potential for condensed timelines in AI-assisted operations.

\section{Assessing Taxonomy Quality}
\label{sec:5-assess-quality}

\cite{taxonomy-dev} note that a useful taxonomy is one that is concise, robust, comprehensive, extendible, and explanatory. These five characteristics constitute the subjective ending conditions in their proposed method for taxonomy development. In our case, the development process proceeded as follows: we first derived the six attributes  from a targeted review of CBRN and offensive cyber threat literature \citep{rand-secure-model-weights, rand-bio-measure, us-cbrn-terrorism-2001, cyber-attack-chain-lockheed} and then went attribute by attribute to distinguish tiers through conceptual or empirical approaches. Finally, we conducted a last iteration reviewing each attribute and its corresponding tier gradient for stability, finding no changes were necessary. The order in which the taxonomy is presented in this paper mirrors this developmental sequence. In Tables \ref{tab:objective-ending-conditions} and \ref{tab:subjective-ending-conditions} below, we assess the eight objective and five subjective ending conditions to evaluate the proposed taxonomy's fitness for purpose. For clarity, we note that \cite{taxonomy-dev}'s ``objects," ``dimensions," and ``characteristics" correspond in our taxonomy to potential frontier AI misuse threat actors, attributes, and tiers respectively.

\begin{longtable}{>{\raggedright\arraybackslash}p{3.5cm}
                  >{\raggedright\arraybackslash}p{10.5cm}}

    \caption{Assessing Objective Ending Conditions}
    \label{tab:objective-ending-conditions} \\
    \toprule
    \textbf{Ending Condition} & \textbf{Our Assessment} \\
    \midrule
    \endfirsthead

    \caption[]{Assessing of Objective Ending Conditions (continued)} \\
    \toprule
    \textbf{Ending Condition} & \textbf{Our Assessment} \\
    \midrule
    \endhead

    \bottomrule
    \multicolumn{2}{r}{\footnotesize\textit{Continued on next page}} \\
    \endfoot

    \bottomrule
    \multicolumn{2}{p{14.0cm}}{%
        \footnotesize\textit{Note.} Objective ending conditions per \cite{taxonomy-dev} ensure the proposed taxonomy has dimensions with mutually exclusive and collectively exhaustive characteristics.} \\
    \endlastfoot


    \textit{All objects or a representative sample of objects have been examined}
        & The taxonomy was developed through a full review of CBRN and cyber threat analysis literature such that the attributes derived came from representative real-world considerations. Each attribute was then divided into tiers based on conceptual or empirical approaches of existing theories and studies on adversaries. \\
    \midrule

    \textit{No object was merged with a similar object or split into multiple objects in the last iteration}
        & The final iterations of tier derivation for each attribute produced stable classifications without requiring further consolidation or subdivision of any possible threat actors along that attribute. \\
    \midrule

    \textit{At least one object is classified under every characteristic of every dimension}
        & Each tier across all six attributes is populated by at least one plausible threat actor type, as explained in the figures in the previous sections. \\
    \midrule

    \textit{No new dimensions or characteristics were added in the last iteration}
        & The six attributes and their respective tiers reached a stable configuration prior to the final iteration, with no additional attributes or tiers needing to be introduced thereafter. \\
    \midrule

    \textit{No dimensions or characteristics were merged or split in the last iteration}
        & The attribute boundaries and tier gradients remained unchanged in the final iteration. \\
    \midrule

    \textit{Every dimension is unique and not repeated}
        & Each of the six attributes captures a conceptually distinct aspect of adversary capacity. \\
    \midrule

    \textit{Every characteristic is unique within its dimension}
        & Within each attribute, tiers are defined by non-overlapping criteria. \\
    \midrule

    \textit{Each cell (combination of characteristics) is unique and is not repeated}
        & Because each attribute is assessed independently and tier assignments do not constrain one another, every combination of tiers across the six attributes produces a unique profile cell. The matrix structure ensures that no two distinct combinations of tiers can describe the same threat actor. \\

\end{longtable}

\begin{longtable}{>{\raggedright\arraybackslash}p{3.5cm}
                  >{\raggedright\arraybackslash}p{10.5cm}}

    \caption{Assessing Subjective Ending Conditions}
    \label{tab:subjective-ending-conditions} \\
    \toprule
    \textbf{Ending Condition} & \textbf{Our Assessment} \\
    \midrule
    \endfirsthead

    \caption[]{Assessing Subjective Ending Conditions (continued)} \\
    \toprule
    \textbf{Ending Condition} &  \textbf{Our Assessment} \\
    \midrule
    \endhead

    \bottomrule
    \multicolumn{2}{r}{\footnotesize\textit{Continued on next page}} \\
    \endfoot

    \bottomrule
    \multicolumn{2}{p{14.0cm}}{%
        \footnotesize\textit{Note.} Subjective ending conditions assessing the proposed taxonomy's usefulness, following \cite{taxonomy-dev}.} \\
    \endlastfoot


    \textit{Concise}
        & The taxonomy contains six dimensions, for a total of 29 tiers across them. The number of dimensions falls within the seven plus or minus two range identified by \cite{seven-processing} as the limit of manageable complexity, and the tier counts within each attribute are similarly bounded, ranging from four to five tiers per attribute. \\
    \midrule

    \textit{Robust}
        & The six attributes collectively capture the structural dimensions along which threat actors meaningfully vary in their capacity to cause harm with AI systems, as justified in each of the ``Why It Matters'' subsections for each attribute. \\
    \midrule

    \textit{Comprehensive}
        & The taxonomy classifies the full range of threat actors currently characterized in frontier safety policies and concerns highlighted in relevant CBRN and offensive cyber literature. The lowest and highest tiers of each attribute are left unbounded. \\
    \midrule

    \textit{Extendible}
        & New attributes can be added as rows and new tiers can be inserted into existing attributes without disrupting the existing matrix structure. The lowest and highest tiers of each attribute are left unbounded. \\
    \midrule

    \textit{Explanatory}
        & The taxonomy does not merely catalogue threat actors but illuminates the structural properties that make them more or less capable of causing harm with AI systems, as grounded via our meta-characteristic choice. As Section~\ref{sec:6-building-complete} highlights, two actors may appear similar if underspecified but will pose meaningfully different risks --- which is precisely why evaluation design must attend to each attribute separately. \\

\end{longtable}

\chapter{Operationalizing Attribute Analysis}
\label{chp:6}

Chapter \ref{chp:5} established the six attributes and their respective tiers in our proposed taxonomy. This chapter turns to their application, first by showing how the attributes can be combined to create complete and precise threat actor profiles, and then by highlighting evaluation design considerations that follow from each attribute. Finally, we discuss how this taxonomy can be integrated into institutional practice. As such, the following sections demonstrate how the proposed taxonomy moves from theory to action, vagueness to clarity: a populated profile is not an end in itself, but rather a set of constraints that can hold evaluations accountable through explicit, pre-specified design choices. 

\section{Building a Complete Threat Actor Profile}
\label{sec:6-building-complete}

 \begin{figure}[!htbp]
     \centering
     \includegraphics[width=0.425\linewidth]{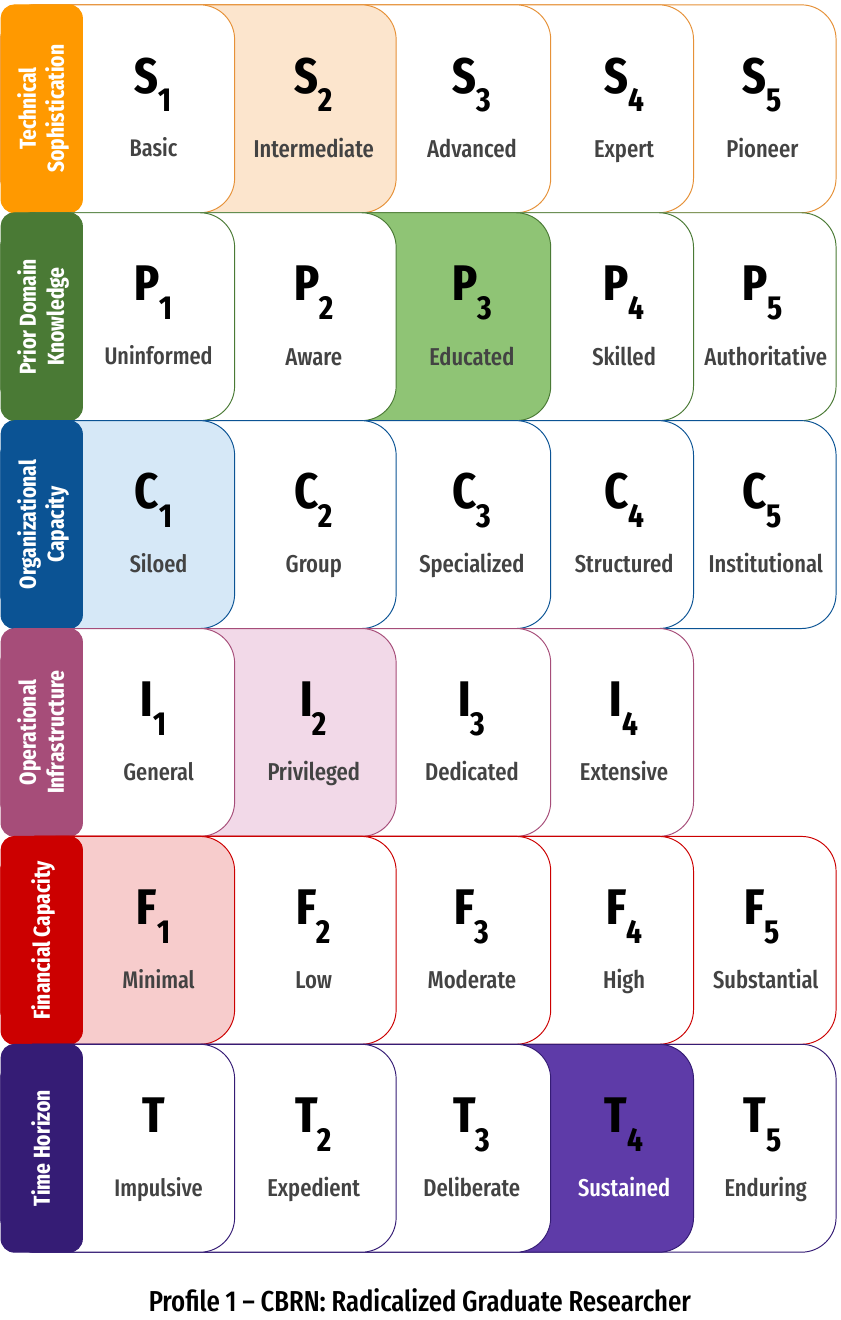}
     \hspace{40pt}
      \includegraphics[width=0.425\linewidth]{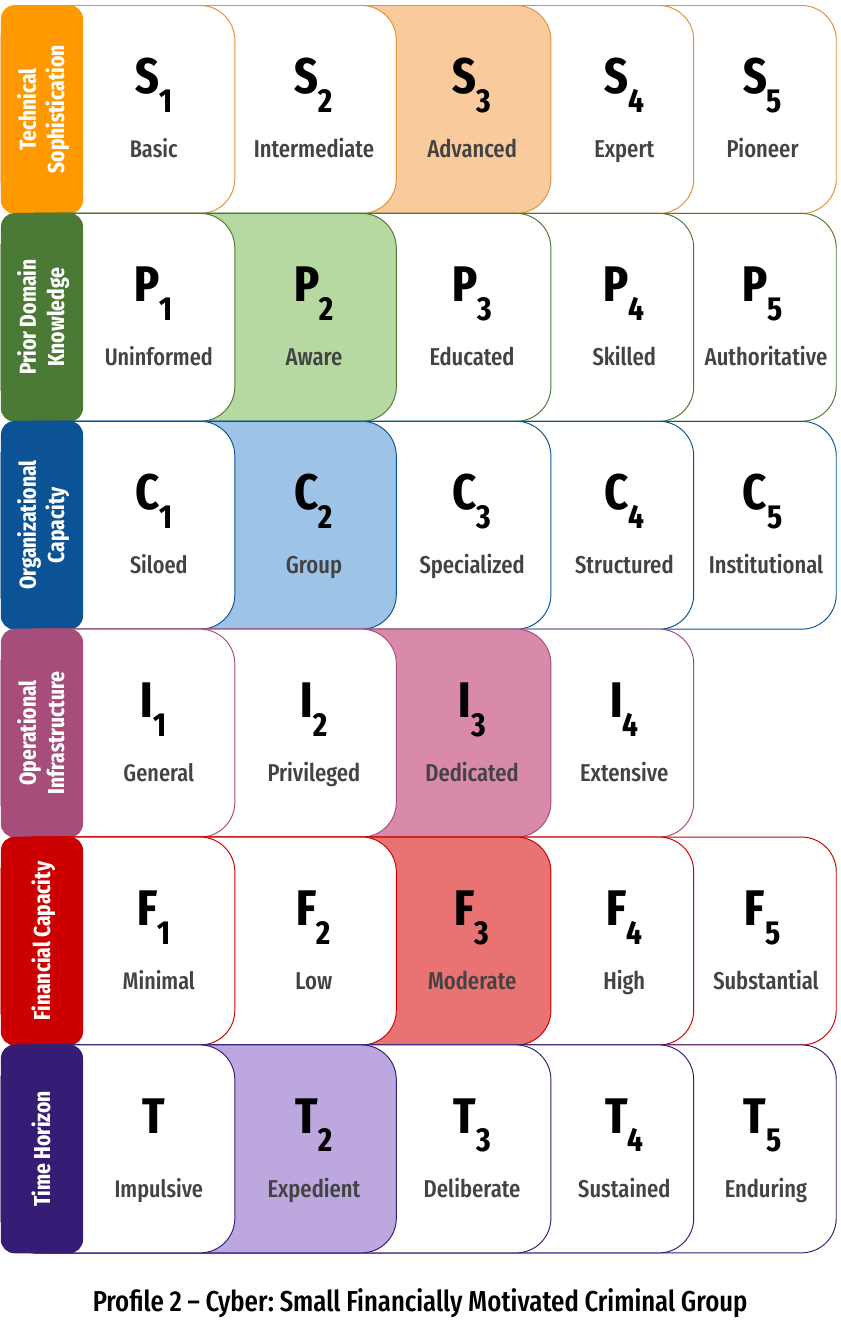}
     \caption[Completed Threat Actor Taxonomy Examples]{Examples of completed threat actor taxonomies.}
     \label{fig:example_profiles}
 \end{figure}

 The taxonomy is most useful when actively populated for a specific adversary of concern rather than left as an abstraction. We present two examples of complete threat actor profiles (one for the CBRN domain and one for offensive cyber) to demonstrate how the matrix can be utilized in practice (see Figure \ref{fig:example_profiles}). While the examples below specify a single tier per attribute, the taxonomy also supports range specifications. For instance, $T_1 - T_3$ references threat actors with time horizons of one month or less. This flexibility can be helpful where genuine uncertainty exist about an actor, or when reasoning about a population of different actors.

 \textbf{Example CBRN Profile: Radicalized Graduate Researcher} 
 
 \textit{Attribute Vector}: [$S_2$, $P_3$, $C_1$, $I_2$, $F_1$, $T_4$]

A radically motivated individual with graduate-level training in microbiology, operating alone with access to a shared university laboratory. They are ideologically committed and have been pursuing their objective over several months, but have limited financial resources and minimal experience with AI systems beyond the basic consumer interfaces. 

This profile is notably jagged: time horizon and prior domain knowledge are the actor's strongest characteristics, while technical sophistication and financial capacity are low. An evaluation implication for such a case could be that AI provides relatively little value as a knowledge gap filler, as they already possess the relevant expertise. A more pertinent question would be whether an AI system provides meaningful assistance as a collaborator or planning aid for someone who generally knows what they want to do but faces logistical and procedural constraints. Perhaps evaluations targeting this profile should therefore prioritize agentic planning tasks over knowledge retrieval benchmark, and use participants with genuine domain expertise (e.g., PhD or postdoctoral fellows) rather than novices (e.g., second-year undergraduates) as a proxy.

\textbf{Example Cyber Profile: Small Financially Motivated Criminal Group} 
 
\textit{Attribute Vector}: [$S_3$, $P_2$, $C_2$, $I_3$, $F_3$, $T_2$]

A small group of five actors motivated by financial gain, operating under informal leadership with access to commercially available exploit tooling and rented infrastructure. They have intermediate technical fluency with AI systems and moderate financial resources, but limited target-specific domain knowledge and no organizational depth beyond their immediate circle. Their time horizon is short, as they optimize for near-term payoff runs than sustained operations. 

Compared with the first example, jaggedness spans in the opposite direction: technical sophistication and financial capacity are moderate, but domain knowledge is more of a constraint. Perhaps most useful to this group is having AI fill in knowledge gaps, providing target-specific reconnaissance, vulnerability context, and operational detail. Evaluations that target this profile should assess whether the model meaningfully speeds up target research and lowers the expertise threshold for executing specific attack methods. For this specific case, a design choice could be employing multiple, continuous sessions that reflect the group's capacity for coordinated but time-bound engagement. 

These two profiles, despite occupying broadly similar positions in terms of overall capability, warrant different evaluation priorities and setups. The first calls for domain sub-expert participants \citep{ouagrham-bioweapons} and agentic planning tasks; the second calls for multi-session designs and knowledge retrieval benchmarks. A characterization that described both simply as ``moderately capable actors" would collapse and obscure those differences. Hence, these profiles illustrate the taxonomy's core utility: not describing actors with underspecified, conflated terms, but making the specific combination of constraints and capabilities visible to calibrate evaluation design against. 
 
\section{Evaluation Design Considerations}
This section does not intend to be exhaustive. For each attribute, we identify an evaluation design consideration that follows directly from engaging in explicit threat actor characterization -- intended to exemplify the kinds of questions our proposed taxonomy surfaces. The goal is to open more structured conversations about how threat actor assumptions can be translated into concrete design choices, which we hope future work builds upon to enable a more rigorous evaluation methodology for frontier AI misuse risks.

\textbf{Technical Sophistication}, \textit{Tiered Elicitation Protocols}. Evaluations should explicitly match elicitation effort to the technical sophistication tier being modeled. A \textit{Basic}-tier evaluation might test prompt-based jailbreaks, while an \textit{Expert}-tier evaluation could include fine-tuning, systematic prompt optimization, and scaffold-augmented workflows. A robust practice would be making this mapping explicit and pre-registered rather than ad hoc, rendering the elicitation strategy a documented consequence of the assumed threat actor profile. 

\textbf{Prior Domain Knowledge}, \textit{Preprepared Subtask Nudging}. Evaluations can operationalize an adversary's assumed knowledge tier by generating domain-specific subtasks that approximate what the actor already knows, presenting them as contextual nudges so that measurement is not confounded by failure at steps the actor would not actually need AI assistance to complete. Rather than presenting the full task from scratch, this approach may better simulate the interaction dynamic of an actor who brings partial capability and uses the model to address a specific gap. 

\textbf{Organizational Capacity}, \textit{Role-differentiated, Sustained Engagement}. Organizational capacity implies having some sort of ability to maintain persistent, distributed engagement toward an objective across personnel. Single-session and single-actor evaluations by design may underrepresent what an organizationally enabled adversary can accomplish. Evaluations seeking to target higher organizational tiers could test under settings of sustained, multi-session engagement with division of labor among participants. 

\textbf{Operational Infrastructure},  \textit{Declared Environment Mapping.} The environment made available to participants (e.g., software libraries, agent scaffolding, physical laboratory equipment) necessarily make implicit assumptions. We recommend that evaluators explicitly justify how their proxy environment maps onto a corresponding tier here in our taxonomy. Doing so not only helps ground the evaluations' assumptions in a declared referent, but also make comparison across studies more faithful, where they would otherwise be confounded by unacknowledged differences in environmental configuration.

\textbf{Financial Capacity}, \textit{Budget-calibrated Scoping}. Large companies of similar scale to frontier AI companies routinely allocate substantial amounts of money to fortify their cybersecurity posture \footnote{In 2021, Microsoft announced a \$20 billion, five-year cybersecurity spending commitment: https://finance.yahoo.com/news/microsoft-commits-to-spend-20-billion-on-cybersecurity-213039278.html.}. A point worth raising is whether capability evaluations for frontier AI misuse risks should be scoped and budgeted with similar seriousness. If the adversaries of concern operate at moderate to high financial capacity tiers, evaluations designed under relatively trivial monetary expenditures may be systematically underestimating what those actors can do. 

\textbf{Time Horizon}, \textit{Accounting for Warmup Time}. At a mechanistic level, the structure of time conducted within an evaluation matters independently of its total duration. Participants are likely unfamiliar with the evaluation environment at the outset, meaning the first moments of engagement may be spent orienting rather than executing. Evaluations that allocate insufficient time relative to the time horizon tier being modeled may underestimate the capabilities of real adversaries -- not because the model is less capable, but because the study design did not allow participants to reach full operational tempo.

\section{Toward Institutional Practice}
Section \ref{sec:6-building-complete} provided examples using the proposed taxonomy to construct complete threat actor profiles. This exercise is ideally one component of a broader system of institutionalized practice. In a way, the taxonomy and accompanying concepts introduced in this paper could be viewed as setting the foundation as the ISO 9000:2015 standard sets the foundation for standardizing quality management systems (QMS)\footnote{The ISO 9000:2015 standard, which we will refer to as ISO 9000 for brevity, is the heart of one of the world's most widely recognized quality management frameworks, which has been implemented by 1.2+ million organizations spanning 180+ countries \citep{iso-9000-library}.}. ISO 9000 outlines seven quality management principles\footnote{These are: customer focus, leadership, engagement of people, process approach, improvement, evidence-based decision making, relationship management \citep{iso-9000-library}.} and definitions for 138 key terms, all to empower effective implementation of quality management systems and clear interoperability; our taxonomy outlines six adversary-relevant attributes and 29 total tiers to establish clarity in describing potential threat actors for frontier AI misuse risks. In each of their respective domains, ISO 9000 and our taxonomy supply the needed conceptual foundation and universal language for downstream use. 

\vspace{8pt}

\begin{figure}[!htbp]
     \centering
     \includegraphics[width=0.6\linewidth]{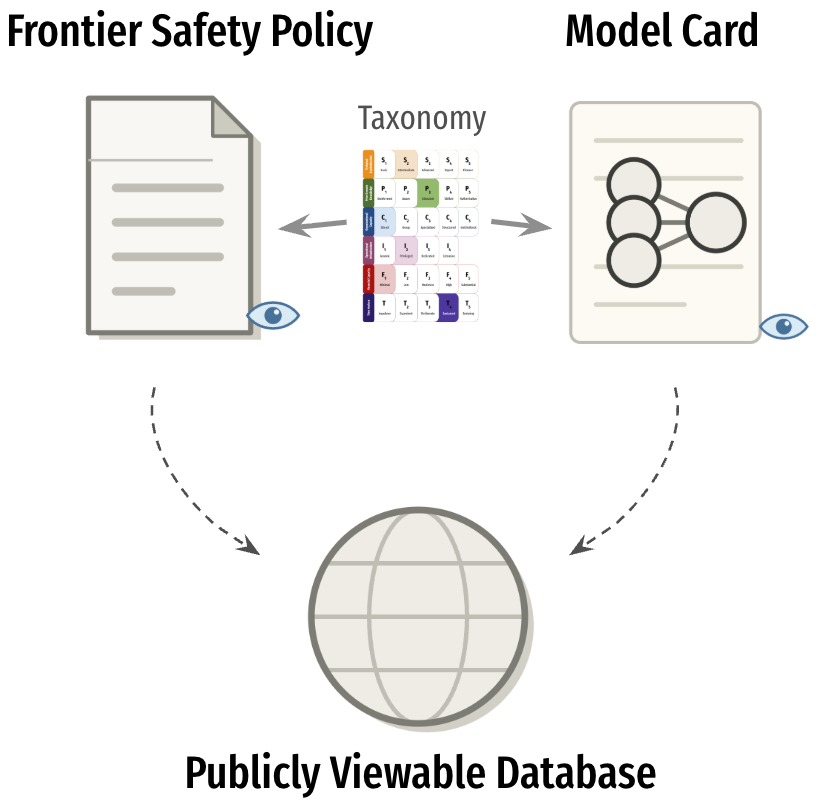}
     \vspace{5pt}
     \caption[Scheme for Institutional Practice]{The taxonomy integrates into two existing practices: frontier safety policies, where developers describe threat actors, and model cards, where evaluation details are reported. With third-party verification, these characterizations could be logged into a publicly viewable database.}
     \label{fig:proposal_figure}
 \end{figure}

In a minimally intrusive manner, we envision the taxonomy fitting into a two-part structure that has emerged in industrial practice and is now being formalized in frontier AI governance. Firstly, frontier AI companies, within their FSPs\footnote{Creating such policies voluntarily saw international traction in May 2024 at the AI Seoul Summit \citep{seoul-commitments}, and some policymakers are now beginning to mandate it (see California SB-53 §22757.12(a); New York RAISE Act §1421).}, should use all six attributes to detail the threat actors they are concerned about and against which their evaluations are calibrated, as exemplified in Section \ref{sec:6-building-complete}. These profiles will serve as the standing policy and can be modified over time. Secondly, in every model/system card that accompanies each frontier model release\footnote{Per transparency measures, revealing model information on evaluation procedures is also beginning to take regulatory shape (see California SB-53 §22757.12(c); New York RAISE Act §1421; EU AI Act Article 53(1)(a) and Annex XI).}, evaluation descriptions should mention how the study attended to each of the six attributes.

For verification purposes, an external independent body could audit frontier AI companies to ensure that (1) taxonomy attributes are sufficiently addressed in both FSP and model card writings and (2) the company conforms to their documented setup. This would be rather similar to the ISO 9000 family system, where organizations seeking ISO 9001 certification for their QMS must go through a third-party audit process \citep{iso-9001-audit}. Complementing this further to assist decision makers, this external body could simultaneously track and file these studies' profiles and their subsequent evaluation results into a publicly viewable registry\footnote{Some dual-use information should be redacted, with only appropriate details made  publicly available.}, organized by the taxonomy matrix to indicate which attributes -- and at what tiers -- have been covered and where gaps remain. This mirrors the logic of pre-analysis plans in medicine and economics, where pre-commitment was introduced precisely because the absence of it produced systematic distortions in reported results; an analogy to this registry for frontier AI capability evaluations could be the American Economic Association (AEA) registry for economic RCTs, where instead threat actor profiles are collected and timestamped before evaluation results are known \citep{minimizing-rct-risks}. Table \ref{tab:iso-tacs-comparison} summarizes this two-part verification structure alongside the ISO 9001 analogue; Figure \ref{fig:proposal_figure} displays the overall vision for institutional practice.

\begin{longtable}{>{\raggedright\arraybackslash}p{2.75cm}
                  >{\raggedright\arraybackslash}p{5.5cm}
                  >{\raggedright\arraybackslash}p{5.5cm}}

    \caption[ISO 9001 Auditing v. Proposed Threat Actor Characterization System]{ISO 9001 Auditing v. Proposed Threat Actor Characterization System (TACS)}
    \label{tab:iso-tacs-comparison} \\
    \toprule
    \textbf{Step} & \textbf{ISO 9001 (Quality Management)} & \textbf{Threat Actor Characterization} \\
    \midrule
    \endfirsthead

    \caption[]{ISO 9001 Auditing v. Proposed TACS (continued)} \\
    \toprule
    \textbf{Step} & \textbf{ISO 9001} & \textbf{Threat Actor Characterization System (TACS)} \\
    \midrule
    \endhead

    \bottomrule
    \multicolumn{3}{r}{\footnotesize\textit{Continued on next page}} \\
    \endfoot

    \bottomrule
    \multicolumn{3}{p{14.5cm}}{%
        \footnotesize\textit{Note.} The proposed TACS does not include a pre-assessment or periodic recertification step. The former is likely unnecessary given that FSPs are public documents and readiness can be assessed without a separate visit. The latter is also likely unnecessary because TACS audits are event-driven rather than time-bounded: every qualifying FSP update triggers a Stage 1 audit and every frontier model release triggers a Stage 2 audit, ensuring continuous compliance verification with no gap in which drift can accumulate. \cite{iso-9001-audit} provides the outline of stages for the ISO 9001 audit process.}\\
    \endlastfoot


    Preparation
        & Prepare a Quality Management System (QMS), including internal audits and management review 
        & Prepare a Frontier Safety Policy (FSP), characterizing potential threat actors with all six taxonomy attributes \\
    \midrule

    (Optional) \newline
        & A pre-assessment or gap-analysis style visit to check readiness 
        & --- \\
    \midrule

    Stage 1 Audit
        & Document review and readiness check 
        & FSP document review, verifying that all six attributes are sufficiently addressed \\
    \midrule

    Stage 2 Audit
        & Full certification assessment of how the organization works in practice 
        & Per-model evaluation and card review, verifying that evaluation procedures conform to the documented threat actor profiles \\
    \midrule

    Ongoing Record
        & Surveillance audits in years one and two of the three-year certification cycle 
        & Stage 1 audits repeat with each FSP update, if threat actor characterizations are modified\vspace{6pt}\newline Stage 2 audits repeat with each frontier model release; evaluation setup and results are stored and publicly displayed\\
    \midrule

    Periodic Renewal
        & Recertification audit in year three 
        & --- \textit{(see note below)} \\

\end{longtable}


\chapter{Conclusion}
\label{chp:7}

This paper proposes a six-attribute taxonomy for characterizing threat actors in pre-release risk management of frontier AI models, addressing the absence of a standardized structure for adversary analysis across the field. By decomposing threat actors into technical sophistication, prior domain knowledge, organizational capacity, operational infrastructure, financial capacity, and time horizon, the taxonomy provides a common vocabulary for articulating who an evaluation is calibrated against, and a basis for holding evaluation design accountable to that specification. The core argument is that for capability evaluations to be reliably interpreted, compared, or improved, there must be explicit adversary assumptions, and that those assumptions are currently made inconsistently and often implicitly across organizations. The practical implication is straightforward: before designing an evaluation, developers should populate a threat actor profile using the taxonomy, and treat that profile as a pre-commitment that constrains subsequent design choices. The taxonomy developed here would serve as the structured vocabulary for such a registry's adversary specification fields.

The application of this paper's proposed approach is particularly urgent for open-weight model developers, who face the most acute pre-release risk management challenge and have, to date, largely not engaged with structured adversary profiling as part of their evaluation practice. Unlike closed API deployments, open-weight releases are irreversible: weights cannot be recalled, safeguards cannot be patched, and usage cannot be monitored. The decision must therefore be made prospectively, which makes the kind of explicit, structured adversary reasoning this taxonomy supports critical.

\bibliography{ref/refs}

\end{document}